\documentclass[fleqn,usenatbib]{mnras}

\usepackage[T1]{fontenc}
\usepackage{ae,aecompl}

\usepackage{graphicx}	
\usepackage{amsmath}	
\usepackage{amssymb}	
\usepackage{bm}
\usepackage{tabularx} 
\usepackage{color}
\usepackage{times}
\usepackage{hyperref}
\usepackage[normalem]{ulem}

\makeatletter
\newcommand{\HI}{{\rm H\,{\scriptstyle I}}}

\newcommand{\Rmnum}[1]{\expandafter\@slowromancap\romannumeral #1@}

\newcommand{\LyA}{\mbox{Ly}\alpha}
\newcommand{\NHI}{N_{\mbox{\tiny H\Rmnum{1}}}}

\newcommand{\CDDF}{\frac{\partial^2\mathcal{N}}{\partial\NHI\partial z}}
\makeatother

\title[Ly$\alpha$ RT in emission and absorption]
{A new model framework for circumgalactic Ly$\alpha$ radiative transfer constrained by galaxy-Ly$\alpha$ forest clustering}

\author[K. Kakiichi and M. Dijkstra] 
{Koki Kakiichi$^{1}$\thanks{E-mail: k.kakiichi@ucl.ac.uk} and
Mark Dijkstra$^{2}$
\\
$^1$ Department of Physics and Astronomy, University College London, London, WC1E 6BT, UK \\
$^2$ Institute of Theoretical Astrophysics, University of Oslo, Postboks 1029 Blindern, N-0315 Oslo, Norway\\
}
\begin{document}
\label{firstpage}
\pagerange{\pageref{firstpage}--\pageref{lastpage}} \pubyear{2017}
\maketitle


\begin{abstract}
We present a new perturbative approach to ``constrained Ly$\alpha$ radiative transfer'' (RT) through the circum- and inter-galactic medium (CGM and IGM). We constrain the $\HI$ content and kinematics of the CGM and IGM in a physically motivated model, using the galaxy-Ly$\alpha$ forest clustering data from spectroscopic galaxy surveys in QSO fields at $z\sim2-3$. This enables us to quantify the impact of the CGM/IGM on Ly$\alpha$ emission in an observationally constrained, realistic cosmological environment. Our model predicts that the CGM and IGM at these redshifts transmit $\approx80~\%$ of Ly$\alpha$ photons after having escaped from galaxies. This implies that while the inter-stellar medium primarily regulates Ly$\alpha$ escape, the CGM has a non-negligible impact on the observed Ly$\alpha$ line properties and the inferred Ly$\alpha$ escape fraction, even at $z\sim 2-3$. Ly$\alpha$ scattering in the CGM and IGM further introduces an environmental dependence in the (apparent) Ly$\alpha$ escape fraction, and the observed population of Ly$\alpha$ emitting galaxies: the CGM/IGM more strongly suppresses direct Ly$\alpha$ emission from galaxies in overdense regions in the Universe, and redistributes this emission into brighter Ly$\alpha$ haloes. The resulting mean surface brightness profile of the Ly$\alpha$ haloes is generally found to be a power-law $\propto r^{-2.4}$. 
Although our model still contains arbitrariness, our results demonstrate how (integral field) spectroscopic surveys of galaxies in QSO fields constrain circumgalactic Ly$\alpha$ RT, and we discuss the potential of these models for studying CGM physics and cosmology.
\end{abstract}


\section{Introduction}

Understanding the interaction between galaxies and the surrounding circum- and inter-galactic medium (CGM and IGM) is key to understanding many outstanding problems in modern extragalactic astrophysics. The role of feedback and environment such as cold streams \citep{2005MNRAS.363....2K,2009Natur.457..451D} and galactic winds \citep{2005ApJ...618..569M,2005ARA&A..43..769V} has been a central theme in regulating the evolution of galaxies throughout cosmic history \citep{1991ApJ...379...52W,2006MNRAS.365...11C,2014MNRAS.445..581H,2015MNRAS.446..521S}. The structure of the CGM and IGM around galaxies are influenced both by supernova and black hole feedback, as well as the large-scale gaseous environment. This makes the CGM an ideal laboratory to test galaxy formation theories. 

The structure and kinematics of $\HI$ gas in the CGM and IGM directly affects observable properties of Ly$\alpha$ line emission. Examples of these properties include the observed Ly$\alpha$ spectral line shape \citep[e.g.][]{Gronke16,Gronke17}, and the surface brightness profiles of extended Ly$\alpha$ haloes around galaxies \citep[e.g.][]{Cantalupo14, 2015Sci...348..779H}. A proper understanding of circumgalactic and intergalactic Ly$\alpha$ radiative transfer (RT) would allow us to use observations of the Ly$\alpha$ emission line into a unique window on the circum-/inter-galatic gaseous environment. 

In addition, understanding circumgalactic/intergalactic Ly$\alpha$ radiative transfer has several cosmological applications: ({\it i}) understanding how the IGM and CGM affect Ly$\alpha$ line emission is crucial when using galaxies at $z>6$ to probe the Epoch of Reionization (e.g. \citealt{2011MNRAS.414.2139D, 2015MNRAS.446..566M,2015MNRAS.452..261C,2016MNRAS.463.4019K,2017ApJ...839...44S,2017arXiv170905356M}, see e.g. \citealt{2014PASA...31...40D} for an extensive review) ({\it ii}) cosmological surveys of Ly$\alpha$ emitting galaxies $z\sim 2-4$ which aim to constrain the dark energy equation of state (e.g. with HETDEX, \citealt{2008ASPC..399..115H}) may be sensitive to Ly$\alpha$ RT effects, as the Ly$\alpha$ RT through the IGM modulates the apparent visibility of Ly$\alpha$ emission line of galaxies (\citealt{2011ApJ...726...38Z},\citealt{2013A&A...556A...5B}, and this paper). The Ly$\alpha$ RT effect in the clustering of Ly$\alpha$ emission has tentatively been detected in BOSS data by \citet{2016MNRAS.457.3541C}, although controversy still remains. Ly$\alpha$ RT effects can therefore affect redshift-space distortions (RSD) in way that can affect the accuracy with which cosmological parameters can be constrained \citep{2011MNRAS.415.3929W}. A combined analysis of the three-point correlation function would be required to circumvent the issue \citep{2013MNRAS.431.1777G}.

 On the theoretical side, modelling Ly$\alpha$ transfer through multiphase interstellar medium (ISM), CGM and IGM from first principles is a daunting task (e.g. \citealt{2015PASA...32...27H}, \citealt{Gronke16,Gronke17} \citealt{McCourt17}, \citealt{Lecture}), partially because existing numerical simulations do not have the resolution to properly resolve the cold, neutral phase \citep{McCourt17}. In addition, it has been difficult to observationally disentangle the impact of the ISM, CGM and IGM on observed Ly$\alpha$ line shapes of galaxies: we currently do not understand to what extent observed Ly$\alpha$ spectral line shapes, and even observationally inferred Ly$\alpha$ escape fractions, are shaped by small scale radiative transfer in the ISM inside galaxies versus the intermediate-to-large scale CGM and IGM outside galaxies. Disentangling the two effects is of great interest for understanding the gaseous environment of galaxies (and therefore galaxy formation), and for being able to use Ly$\alpha$ emitting galaxy surveys for cosmological purposes. 

 Spectroscopic surveys of galaxies in the foregrounds of quasars (QSOs) provide a potentially unique observational window on Ly$\alpha$ radiative transfer on a range of scales, as they enable us to study both the ISM properties of galaxies from direct spectroscopy, and the surrounding CGM and IGM from the Ly$\alpha$ absorption in the background QSO spectra (at $z\sim2-3$, see \citealt{2003ApJ...584...45A,2005ApJ...629..636A,2006ApJ...652..994C,2011MNRAS.414...28C,2012ApJ...751...94R,2012ApJ...750...67R,2014MNRAS.442.2094T,2014MNRAS.445..794T}).  Despite the enormous scientific potential of this approach (\citealt{2010ApJ...717..289S,2011ApJ...736..160S,2013ApJ...766...58H,2015Sci...348..779H}), a theoretical framework that is needed to fully exploit these observations is still in its infancy. Most previous works investigating the connection between galaxies/quasars and their gaseous environments focused either on Ly$\alpha$ emission \citep{2005ApJ...628...61C,2010ApJ...725..633F,2010ApJ...708.1048K,2015ApJ...806...46L,2016ApJ...822...84M} or Ly$\alpha$ absorption around galaxies and quasars \citep{2011MNRAS.418.1796F,2013ApJ...765...89S,2015MNRAS.452.2034R, 2015MNRAS.453..899M,2016MNRAS.461L..32F}. \citet{2012MNRAS.424.1672D} introduced ``constrained Ly$\alpha$ radiative transfer'' in which Ly$\alpha$ circumgalactic RT modelling was performed through models of the CGM in which its $\HI$ content was constrained by the Ly$\alpha$ absorption signal in the background galaxies and QSOs (see also \citealt{2012A&A...540A..63N,2013MNRAS.433.3091K,2017ApJ...837...71C}). This approach enables controlled numerical experiments of Ly$\alpha$ radiative transfer through physical realistic distributions of $\HI$ in the CGM/IGM.
 
In this paper, we present an improved constrained Ly$\alpha$ RT model of the impact of the CGM and IGM on Ly$\alpha$ emission in and around galaxies. Earlier work on circumgalactic/intergalactic Ly$\alpha$ RT either employed analytic models for the density and kinematics of $\HI$ gas in the CGM/IGM (\citealt{2007MNRAS.377.1175D}, also see \citealt{2004MNRAS.349.1137S}) or cosmological (hydro)simulations \citep{2008MNRAS.391...63I,2010ApJ...716..574Z,2011MNRAS.410..830D,2011ApJ...728...52L}, which - as mentioned above - currently do not resolve the cold, neutral phase of the CGM/IGM (see \citealt{McCourt17} for extended discussion). \citet{2012MNRAS.424.1672D} and \citet{2013ApJ...766...58H} employed a joint Ly$\alpha$ emission - absorption approach to circumgalactic Ly$\alpha$ RT. Their models, none the less, paid attention mostly to the neutral gas {\it distribution} in the CGM and Ly$\alpha$ haloes; the absorption data was not used optimally to constrain the gas {\it kinematics}. This is important as gas kinematics is well known to play an important role in Ly$\alpha$ transfer process. Our model improves upon previous works in two important aspects: ({\it i}) the distribution and kinematics of neutral gas around galaxies are constrained by observations of the galaxy-Ly$\alpha$ forest clustering and its RSD, which provide direct observational constraints on both the $\HI$ distribution and kinematics in the CGM/IGM; ({\it ii}) we model both the impact of the CGM/IGM on the emerging Ly$\alpha$ flux from galaxies (and its spectral line shape), and the predict surface brightness profiles of the scattered Ly$\alpha$ radiation. The latter is important as Ly$\alpha$ scattering simultaneously attenuates direct Ly$\alpha$ emission from galaxies, and gives rise to spatially extended Ly$\alpha$ haloes. Thus, the two observables combined provide a more complete picture of how the CGM and IGM affect Ly$\alpha$ radiation in and around galaxies.

Our aim of this paper is two-fold. The {\it first} is to introduce a new unified statistical, perturbative modelling of Ly$\alpha$ RT and galaxy-Ly$\alpha$ forest clustering in order to achieve the goals ({\it i}) and ({\it ii}). Our model naturally includes absorbers with $\HI$ column densities in the range $\log_{10} N_{\rm HI}/{\rm cm}^{-2}\sim 13-22$. This is important, as the nature of self-shielding absorbers is still not well understood. Moreover, the impact of low column density absorbers surrounding galaxies on the emerging Ly$\alpha$ emission line has not been explored at all yet. The {\it second} is to demonstrate how joint Ly$\alpha$ emission-absorption data can be used to obtain new insights into how Ly$\alpha$ radiation escapes from galaxies, how it subsequently propagates through the CGM/IGM, and how we can use this knowledge to obtain constraints on the physical properties of the CGM/IGM. 

This paper is organised as follows. In \S~\ref{sec:methodology} we summarise the methodology and the general idea behind our work.  In \S~\ref{sec:theory} we introduce a unified approach to cosmological Ly$\alpha$ RT connecting Ly$\alpha$ emission and absorption. In \S~\ref{sec:model} we describe a physical model of galaxies and the gaseous environments used to analyse the data. The joint analysis of the redshift-space anisotropic galaxy-Ly$\alpha$ forest clustering with the Ly$\alpha$ line profiles and Ly$\alpha$ haloes is presented in \S~\ref{sec:analysis}. In \S~\ref{sec:implication} we discuss the implications for the physics of Ly$\alpha$ escape and Ly$\alpha$ haloes in the CGM and cosmology. Our conclusions are summarised in \S~\ref{sec:conclusion}.
Throughout this paper we adopt the flat $\Lambda$CDM cosmology with $\Omega_m=0.3$, $\Omega_\Lambda=0.7$, $h=0.7$. We denote pkpc and pMpc (ckpc and cMpc) to indicate distances in proper (comoving) units.

\begin{figure*}
  \centering
  \includegraphics[angle=0,width=\textwidth]{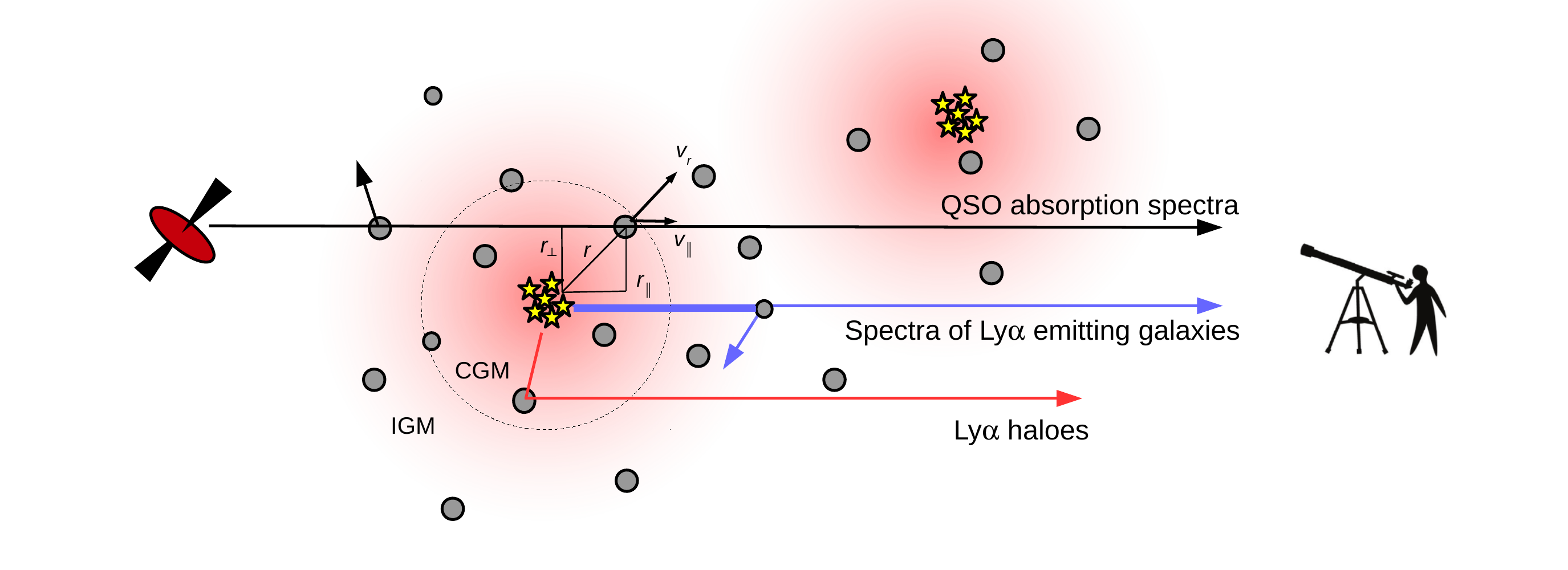}
  \vspace{-0.9cm}
  \caption{Schematic illustration of a spectroscopic survey of Ly$\alpha$ emitting galaxies in QSO fields. Galaxies (represented as a collection of star symbols) reside in gaseous environments composed of the neutral gas (represented by grey-coloured clumps). The dashed circle indicates the viral radius of host dark matter halo. $r,~r_\perp,~r_\parallel$ are the radial, perpendicular, and line-of-sight distance to a gas clump. $v_r$ and $v_\parallel$ are the radial and line-of-sight peculiar velocity of a gas clump relative to a galaxy of interest (see also text). Some Ly$\alpha$ photons emitted from the galaxies are scattered back to observers, contributing to the Ly$\alpha$ haloes (represented by fuzzy red spheres). The cartoon illustrates how ({\it i}) QSO absorption spectra, ({\it ii}) spectra of Ly$\alpha$ emitting galaxies, and ({\it iii}) Ly$\alpha$ haloes around galaxies trace the circum- and inter-galactic gas and their interconnections. }
  \label{fig:cartoon}
\end{figure*}

\section{Methodology: \newline constrained radiative transfer}\label{sec:methodology}

In this section, we describe how our general methodology -- ``constrained radiative transfer'' -- works. A goal is to estimate the impact of the CGM and IGM around galaxies on the Ly$\alpha$ escape fraction and Ly$\alpha$ haloes consistently with the galaxy-Ly$\alpha$ forest clustering measurements (\S~\ref{sec:analysis}). We will discuss the result in the context of the physics of the galaxy-IGM connection and cosmology (\S~\ref{sec:implication}).

Figure~\ref{fig:cartoon} schematically illustrates how the joint analysis of Ly$\alpha$ emission and absorption from a (integral field) spectroscopic survey of galaxies in QSO fields can be used to study the origin of escape fraction and Ly$\alpha$ haloes. This is the most important picture which we base our argument on so that it should be kept in mind throughout the paper. 

First, using galaxy-QSO pairs (black arrow), the galaxy-Ly$\alpha$ forest cross-correlation and its redshift-space distortion can be measured. This can be presented either as a galaxy-transmitted Ly$\alpha$ flux correlation function (e.g. \citealt{2014MNRAS.442.2094T}) or a 2D pixel optical depth map (e.g. \citealt{2014MNRAS.445..794T}). This contains a full statistical information of the average gas structure (grey clumps) around the galaxies both about the gas distribution and velocity field. Thus, by comparing the theoretical and observed Ly$\alpha$ absorption around galaxies, we can fully constrain the average properties of the CGM and IGM around galaxies, in which the Ly$\alpha$ emission from the galaxies propagates through. 

We then perform Ly$\alpha$ RT calculations in the observationally-calibrated medium. As some Ly$\alpha$ photons emitted from galaxies observed in their galaxy spectra (blue arrow) are scattered out of the lines-of-sight by the same CGM and IGM traced by the galaxy-Ly$\alpha$ forest clustering, by performing the Ly$\alpha$ transfer calculation in the medium pre-constrained by the Ly$\alpha$ absorption data, we can estimate the impact of the CGM/IGM on the average Ly$\alpha$ line flux in galaxy spectra, {\it without} adjusting free parameters about the CGM and IGM. Therefore, this enables us to uniquely estimate the contribution of the CGM/IGM on the Ly$\alpha$ escape fraction. As this constrained RT technique substantially reduces the range of allowed parameter space of the model, the calculated impact of the CGM/IGM on the escape fraction is more robust than a conventional approach which does not take into account the galaxy-Ly$\alpha$ forest clustering observations.\footnote{Of course, the parameters about the ISM remain as unconstrained free parameters. However, for computing the escape fraction we are only interested in the fraction of photons attenuated by the CGM/IGM. Thus, it is independent of the values of Ly$\alpha$ luminosity. We are aware that taking into account the detailed ISM Ly$\alpha$ line profile is more difficult. The difficulty, however, persists in virtually all hydrodynamical simulations as our current computational power fails to resolve star-forming clouds, cold gas in the ISM, the multi-phase structures of the CGM and IGM simultaneously in a single cosmological box. We bypass this difficulty and consolidate our predictions by exploring a wide possible range of ISM line profiles.} 

Furthermore, as the Ly$\alpha$ photons (red arrow) that are back-scattered toward an observer will contribute the surface brightness of Ly$\alpha$ haloes (fuzzy red spheres), using the same constrained Ly$\alpha$ RT calculation we can predict the structure of Ly$\alpha$ haloes self-consistently with the mechanism controlling the escape of Ly$\alpha$ photons through the CGM/IGM. Because the observation of Ly$\alpha$ haloes was not used during the calibration step, we can regard it as an independent test of how Ly$\alpha$ radiation emerges in and around galaxies. Thus, being able to reproduce the observed Ly$\alpha$ haloes supports the proposed mechanism and impact of the CGM/IGM on the Ly$\alpha$ escape fraction. In addition, it provides an insight into how the observed structures of the CGM and IGM form Ly$\alpha$ haloes by a proposed mechanism, here, via scattering. 

This methodology -- constrained radiative transfer technique -- uses Ly$\alpha$ absorption and emission simultaneously, which forms our basis for studying the origin of Ly$\alpha$ escape and haloes. In the following section, we will mathematically formulate this argument. We use an analytical model in applying the above methodology. We emphasise, however, that the general methodology is independent of the particular model used. For example, one can employ an analogous analysis using cosmological hydrodynamic simulations of galaxies and Ly$\alpha$ forest, combined with Monte-Carlo Ly$\alpha$ RT simulations.

\section{Ly$\alpha$ radiative transfer}\label{sec:theory}

We introduce a statistical perturbative approach to cosmological Ly$\alpha$ radiative transfer that unifies Ly$\alpha$ in absorption and emission; it connects the physical properties of Ly$\alpha$ forests and self-shielded absorbers with the Ly$\alpha$ emission line profiles of galaxies and Ly$\alpha$ haloes. 

Our principle idea is to formulate the problem of cosmological Ly$\alpha$ radiative transfer {\it using a statistical perturbative approach}. In the context of Ly$\alpha$ absorption, this approach was formulated by \citet{1980ApJ...240..387P} and used in \citet{1996ApJ...461...20H,2012ApJ...746..125H}. The method can be extended to Ly$\alpha$ in emission. Ly$\alpha$ photons undergo multiple scattering; but the contribution from higher order scatterings asymptotically decrease because in each propagation the probability that the photons absorbed and re-emitted decreases as $\propto(1-e^{-\tau_{\rm CGM/IGM}})^{N_{\rm scat}}$ where $N_{\rm scat}$ is the number of Ly$\alpha$ scatterings and $\tau_{\rm CGM/IGM}$ is the optical depth of Ly$\alpha$ photons in a single propagation through the CGM and IGM. This allows us to consider the Ly$\alpha$ transfer process in terms of {\it the perturbative expansion with respect to the number of scatterings}. Ly$\alpha$ photons experiencing zero or single scattering contribute to the largest to the emergent Ly$\alpha$ profile. In {\it the zeroth-order scattering expansion}, we include the Ly$\alpha$ photons that have experienced no scattering; once a photon is scattered out of a line-of-sight, the photon is lost. Thus, Ly$\alpha$ photons are attenuated by $e^{-\tau_{\rm CGM/IGM}}$. The zeroth-order scattering includes the Ly$\alpha$ absorption in the background QSOs and the CGM/IGM attenuation effect on the galaxy spectra (see Figure~\ref{fig:cartoon}).  In {\it the first-order scattering expansion}, we include the contribution from Ly$\alpha$ photons that have scattered back into the line-of-sight after the first scattering. This allows us to estimate the Ly$\alpha$ haloes (see Figure~\ref{fig:cartoon}). In the higher-order scattering expansion, we can extend this to include the contributions from the photons that experienced multiple scatterings, which will approach full Monte-Carlo Ly$\alpha$ RT simulations. 

Therefore, the perturbative approach enables us to self-consistently formulate the Ly$\alpha$ transfer problem both in absorption and emission, connecting the galaxy-Ly$\alpha$ forest clustering (\S \ref{sec:lya_absorption}) with the Ly$\alpha$ emission line profiles of galaxy spectra ({\it the zeroth order scattering expansion}) (\S \ref{sec:spectra}) and Ly$\alpha$ haloes around galaxies ({\it the first order scattering expansion}) (\S \ref{sec:lya_halo}).

\subsection{Ly$\alpha$ absorption around galaxies}\label{sec:lya_absorption}

The Ly$\alpha$ absorption arises from both residual netural gas in the photoionized IGM and self-shielded gas. The Ly$\alpha$ absorption features in QSO spectra are a natural consequence of the continuous cosmic web of the large-scale structure in $\Lambda$CDM cosmology (e.g. \citealt{1994ApJ...437L...9C,2015MNRAS.446.3697L}). In this paper, in order to make the modelling analytically tractable, we approximate the CGM and IGM as consisting of a collection of Lagrangian gas parcels which are referred to as absorbers. 

We define each absorber's velocity and position relative to a galaxy of interest. An absorber located at a comoving distance $r$ away from the associate galaxy has its own $\HI$ column density $\NHI$, temperature $T$, and a proper peculiar velocity $v_r$ relative to the galaxy (see Figure~\ref{fig:cartoon}). The total line-of-sight velocity of an absorber relative to the associated galaxy is $u_\parallel=H(z)r_\parallel/(1+z)+v_\parallel$ where $v_\parallel=\mu v_r$ is the line-of-sight peculiar velocity, $\mu=r_\parallel/r$, and $r_\parallel$ is the comoving distance along the line-of-sight. $H(z)$ is the Hubble parameter at redshift $z$. Thus, for the photons emitted at frequency $\nu_e$ from a galaxy or background QSO, the Ly$\alpha$ optical depth of an absorber is given by
\begin{equation}
\tau_{\rm a}(\nu_e|u_\parallel,\NHI)=\sigma_\alpha\NHI\varphi_\nu\left[T,\nu_e\left(1-\frac{u_\parallel}{c}\right)\right],\label{eq:absorber}
\end{equation}
where $\sigma_\alpha=(\pi e^2/m_ec)f_{12}=0.011\rm~cm^2~Hz$ is the $\LyA$ cross section and $\varphi_\nu(T,\nu)$ is the Voigt profile at gas temperature $T$ and frequency $\nu$, $e$ is the electron charge, $m_e$ is the mass of electron, and $c$ is the speed of light, and $f_{12}=0.4164$ is the oscillator strength of $2P\rightarrow1S$ transition \citep{2009JPCRD..38..565W}. In the following sections, we interchangeably express the emitted frequency $\nu_e$ at the rest-frame of galaxies in terms of the velocity unit $\Delta v$ defined as $\nu_e=\nu_\alpha(1-\Delta v/c)$ and the proper line-of-sight redshift-space coordinate $s_\parallel$ defined as $s_\parallel=\Delta v/H(z)$ (in pMpc) where $\nu_\alpha$ is the frequency at the Ly$\alpha$ line centre.

\subsubsection{The redshift-space anisotropic galaxy-Ly$\alpha$ forest clustering}

The cross-correlation of the transmitted Ly$\alpha$ flux in the Ly$\alpha$ forest region of the background QSO spectra with galaxies' positions traces the statistical properties of the gaseous environments around galaxies. The two-dimensional  effective optical depth map, $\tau_{\rm eff}(\nu_e,r_\perp)$, is defined in terms of the mean transmitted Ly$\alpha$ forest flux around galaxies, $F(\nu_e,r_\perp)$, as a function of emitted frequency, $\nu_e$, at the rest-frame of the galaxies and impact parameter, $r_\perp$, relative to the background QSO (the corresponding proper redshift-space coordinates are $s_\parallel=(c/H)(1-\nu_e/\nu_\alpha)$ and $s_\perp=r_\perp/(1+z)$): 
\begin{equation}
\tau_{\rm eff}(\nu_e,r_\perp)\equiv-\ln\langle F(\nu_e,r_\perp)\rangle.
\end{equation}

As a solution to the statistical perturbative approach to cosmological Ly$\alpha$ transfer (\citealt{2016MNRAS.463.4019K}, see also \citealt{1980ApJ...240..387P,1996ApJ...461...20H}), we find that the 2D effective optical depth map is given by
\begin{align}
\tau_{\rm{eff}}(\nu_e,r_\perp)=&\int d\NHI\frac{\partial^2\mathcal{N}}{\partial\NHI\partial z}\left|\frac{dz}{dr}\right| \label{eq:eff_optdpt}
 \\
&\times\int \frac{du_\parallel}{aH}\left[1+\xi_v(u_\parallel,r_{\perp})\right]\left[1-e^{-\tau_{\rm a}(\nu_e|u_\parallel,\NHI)}\right]\nonumber,
\end{align}
where $\CDDF$ is the $\HI$ column density distribution function (CDDF) of absorbers, $|dr/dz|=c/H(z)$ is the comoving distance per redshift. The velocity-space correlation function between galaxies and absorbers, $\xi_v(u_\parallel,r_{\perp})$, in the Gaussian streaming model\footnote{\citet{1995ApJ...448..494F,2011MNRAS.417.1913R} have shown that the Gaussian streaming model gives the same result as linear theory including Kaiser effect provided that the mean radial velocity and velocity dispersion are computed from linear theory. Because we are interested in the gas flow around galaxies on scales of a few Mpc we do not impose the linear theory. Instead we allows a phenomenological parametrisation of the velocity field.} is given by
\begin{align}
&1+\xi_v(u_\parallel,r_\perp)= \label{eq:velocity_correlation_function} \\
&\int_{-\infty}^{\infty}\frac{aHdr_\parallel}{\sqrt{2\pi\sigma_v^2(r)}}\left[1+\xi(r)\right]\exp\left[-\frac{(u_{\parallel}-aHr_{\parallel}-\mu\langle v_r(r)\rangle)^2}
{2\sigma_{v}^2(r)}\right],\nonumber 
\end{align}
where $\xi(r)$ is the real-space correlation function between galaxies and absorbers, $\langle v_r(r)\rangle$ and $\sigma_v(r)$ are the mean radial velocity and velocity dispersion between galaxies and absorbers at a separation $r$, $r=\sqrt{r_{\parallel}^2+r_{\perp}^2}$, and $a=1/(1+z)$. For simplicity, we assume isotropic velocity dispersion so that both line-of-sight and radial components are the same.

The velocity-space correlation function takes into account the dynamics and clustering of absorbers around galaxies. While we use the Gaussian streaming model as an explicit example, the method can easily be generalized for any pairwise velocity probability distribution function. We model the explicit forms of real-space correlation function and the pairwise mean velocity and velocity dispersion in \S \ref{sec:model}.

\subsubsection{The equivalent width of Ly$\alpha$ absorption}
The rest-frame equivalent width is computed using the same method as \citet{2012ApJ...751...94R}, which measures the excess absorption relative to the mean IGM. The mean equivalent width of Ly$\alpha$ absorption around galaxies is then computed as
\begin{equation}
\langle{\rm EW}(r_\perp) \rangle=\lambda_\alpha\int \frac{d\nu_e}{\nu_\alpha} \left[1-\frac{\langle F(\nu_e,r_\perp)\rangle}{\bar{F}(z)}\right],
\end{equation}
where $\bar{F}(z)=e^{-\bar{\tau}_{\rm eff}(z)}$ and $\bar{\tau}_{\rm eff}(z)$ is the mean effective optical depth of \citet{2013MNRAS.430.2067B}. In the case of $\xi_v=0$, Equation (\ref{eq:eff_optdpt}) reproduces the mean effective optical depth.

\subsection{Ly$\alpha$ emission line profiles of galaxies}\label{sec:spectra}

The Ly$\alpha$ emission line profiles of galaxies are affected by both the ISM-scale RT and the large-scale CGM/IGM environment. We therefore split the model into the small-scale RT part (\S \ref{sec:small}), which includes the multiple scatterings, and the large-scale RT part (\S \ref{sec:large}), which is modelled self-consistently to the Ly$\alpha$ absorption and diffuse Ly$\alpha$ haloes around galaxies using the perturbative expansion.

\subsubsection{Star formation and ISM of galaxies}\label{sec:small}

For Ly$\alpha$ emission due to the nebular recombination in star-forming regions, the average Ly$\alpha$ luminosity is given by 
\begin{equation}
\langle L_\alpha^{\rm intr}\rangle=1.1\times10^{42}\left(\langle{\rm SFR}\rangle/{\rm M_\odot~yr^{-1}}\right){\rm erg~s^{-1}},
\end{equation}
by converting the relation of H$\alpha$ luminosity and star formation rate (SFR) \citep{1998ApJ...498..541K} using the case B approximation \citep{1971MNRAS.153..471B}. We assume the average SFR of galaxies is $\langle{\rm SFR}\rangle=34~{\rm M_\odot~yr^{-1}}$, which is taken from the median dust-corrected SFR of LBGs in \citet{2011ApJ...736..160S}. Therefore, our fiducial value is $\langle L_\alpha^{\rm intr}\rangle=3.7\times10^{43}{\rm~erg~s^{-1}}$.

We assume dust-free multi-phase clumpy ISM. We define the ISM Ly$\alpha$ escape fraction, $\langle f_{\rm esc,ISM}^{\rm Ly\alpha}\rangle$, as a fraction of Ly$\alpha$ photons escaped from the sites of nebular Ly$\alpha$ emission out of galaxies and toward the line-of-sight of an observer. We assume a fiducial value of $\langle f_{\rm esc,ISM}^{\rm Ly\alpha}\rangle=0.20$. Note that in a dust-free ISM only the Ly$\alpha$ scattering by the $\HI$ gas contributes to the ISM escape fraction by scattering off Ly$\alpha$ photons out of the line-of-sight of an observer. Thus in the absence of the CGM and IGM, the observed Ly$\alpha$ luminosity is $\langle f_{\rm esc,ISM}^{\rm Ly\alpha}\rangle\langle L_\alpha^{\rm intr}\rangle$. However, because these scattered photons eventually leak out of the ISM, the photons escaping out of the ISM in {\it all directions} around a galaxy are the same as the intrinsic nebular production of Ly$\alpha$ photons. In other words, the amount of Ly$\alpha$ photons injected into the CGM is the same as $\langle L_\alpha^{\rm intr}\rangle$ in a dust-free ISM model. 

For the intrinsic average Ly$\alpha$ line profile, $\langle\Phi_\alpha^{\rm ISM}(\nu_e)\rangle$, (i.e. emerging from the ISM), we use a result from the Monte-Carlo Ly$\alpha$ RT model of the multi-phase clumpy ISM of \citet{2016ApJ...826...14G} with the $\HI$ column density of $\log_{10}(\NHI/{\rm cm^{-2}})=19.64$ and outflow velocity of $202.2\rm~km~s^{-1}$ (one presented in Figure 6 of \citealt{2016ApJ...826...14G}). This model produces an asymmetric single peak line profile with the Ly$\alpha$ velocity offset of $\langle\Delta v_{\rm Ly\alpha}\rangle=250\rm~km~s^{-1}$. 

Both the ionizing and non-ionizing UV luminosity incorporate results from stellar population synthesis. The average Lyman continuum (LyC, $<912$~\AA) photon production rate (in units of $\rm s^{-1}$) is given by assuming the following conversion from the UV (1500~\AA) luminosity \citep{2013ApJ...768...71R},
\begin{align}
\langle\dot{N}_{\rm ion}\rangle&=\langle f_{\rm esc}^{\rm LyC}\rangle \langle \xi_{\rm ion}\rangle  \langle L_{\rm UV}\rangle,\nonumber \\
&\approx8.6\times10^{52}\left(\frac{\langle f_{\rm esc}^{\rm LyC}\rangle}{0.02}\right)\left(\frac{\langle{\rm SFR}\rangle}{34~\rm M_{\odot}yr^{-1}}\right)\rm~s^{-1},\label{eq:dNiondt}
\end{align}
where we used
\begin{equation}
\langle L_{\rm UV}\rangle=8\times10^{27}\left(\langle\rm SFR\rangle/{\rm M_\odot~yr^{-1}}\right){\rm erg~s^{-1}~Hz^{-1}},
\end{equation}
with the assumptions of a Salpeter IMF and solar metallicity \citep{1998ApJ...498..106M}, and $\rm log_{10}\langle \xi_{\rm ion}\rangle/{\rm erg~Hz^{-1}}=25.2$, consistent with stellar population synthesis model \citep{2013ApJ...768...71R} and observations \citep{2016ApJ...831..176B}. We assume an average LyC escape fraction of $\langle f_{\rm esc}^{\rm LyC}\rangle=0.02$, which is broadly in agreement with the observationally inferred values of the LyC escape fractions (e.g. \citealt{2015ApJ...804...17S,2016A&A...585A..48G}) at $z\sim3$. In this model, the average intrinsic equivalent width of Ly$\alpha$ emission line is $\langle W_{\rm Ly\alpha}^{\rm intr}\rangle=103$~\AA, assuming  the UV slope of $-2$ at $z\sim3$ \citep{2014ApJ...793..115B}.

\subsubsection{Emergent Ly$\alpha$ line profile from CGM/IGM}\label{sec:large}

Ly$\alpha$ photons escaping from the ISM are scattered out of a line-of-sight by the intervening CGM and IGM. Thus, the emergent average Ly$\alpha$ line profile of Ly$\alpha$ emitting galaxies is affected by the opacity of the circum- and inter-galactic gaseous environments. This effect has been formulated by \citet{2016MNRAS.463.4019K} and the CGM/IGM attenuates the average Ly$\alpha$ flux of galaxies as
\begin{equation}
\langle L_\nu(\nu_e)\rangle=e^{-\tau_{\rm eff}^{\rm Ly\alpha}(\nu_e)}\langle f_{\rm esc,ISM}^{\rm Ly\alpha}\rangle\langle L_\alpha^{\rm intr}\rangle \langle\Phi_\alpha^{\rm ISM}(\nu_e)\rangle,
\end{equation}
where $\langle L_\nu(\nu_e)\rangle$ is the apparent specific Ly$\alpha$ luminosity seen by an observer and $\tau_{\rm eff}^{\rm Ly\alpha}(\nu_e)$ is effective optical depth against Ly$\alpha$ line at the emitted frequency $\nu_e$ in the rest-frame of galaxies. The mean transmissivity of the Ly$\alpha$ flux of galaxies is defined as
\begin{equation}
\langle\mathcal{T}_\alpha\rangle=\int e^{-\tau^{\rm Ly\alpha}_{\rm eff}(\nu_e)}\langle\Phi_\alpha^{\rm ISM}(\nu_e)\rangle d\nu_e.
\end{equation}
The observed Ly$\alpha$ EW of the emission line is then given by $\langle W_{\rm Ly\alpha}\rangle=\langle f_{\rm esc,ISM}^{\rm Ly\alpha}\rangle\langle\mathcal{T}_\alpha\rangle\langle W_{\rm Ly\alpha}^{\rm intr}\rangle$.

The CGM/IGM attenuation is caused by the same neutral gas that gives rise to Ly$\alpha$ absorption around galaxies observed in the background QSO spectra. Therefore, our modelling leads the effective optical depth against Ly$\alpha$ line along each sightline of galaxies as,
\begin{align}
\tau_{\rm{eff}}^{\rm Ly\alpha}(\nu_e)=&\int d\NHI\frac{\partial^2\mathcal{N}}{\partial\NHI\partial z}\left|\frac{dz}{dr}\right|\label{eq:eff_optdpt_line}\\
&\times\int \frac{du_\parallel}{aH}\left[1+\zeta_v(u_\parallel)\right]\left[1-e^{-\tau_{\rm a}(\nu_e|u_\parallel,\NHI)}\right].\nonumber
\end{align}
Clearly the CGM/IGM attenuation of the Ly$\alpha$ emission line of galaxies is closely related to the effective optical depth for the galaxy-Ly$\alpha$ absorption clustering (Equation \ref{eq:eff_optdpt}). The difference is that the former is affected only by the foreground gas of galaxies, whereas the latter must also include the gas background of the galaxies. Therefore, the velocity-space correlation function between galaxies and absorbers, $\zeta_v(u_\parallel)$, takes the integral from the innermost radius $r_{\rm min}$ away from a galaxy to an observer at infinity,
\begin{align}
1+\zeta_v(u_\parallel)&=\int_{r_{\rm min}}^{\infty} \frac{aHdr_{\parallel}}{\sqrt{2\pi\sigma_v^2(r_\parallel)}} \label{eq:line_xi} \\
&\times\left[1+\xi(r_{\parallel})\right]\exp\left[-\frac{(u_{\parallel}-aHr_{\parallel}-\langle v_r(r_\parallel)\rangle)^2}
{2\sigma_{v}^2(r_{\parallel})}\right].\nonumber
\end{align}
The real-space correlation function $\xi(r_\parallel)$, the mean radial velocity field $\langle v_r(r_\parallel)\rangle$ and velocity dispersion $\sigma_{v}(r_{\parallel})$ are the same as ones in Equation (\ref{eq:velocity_correlation_function}), but evaluated at $r_\parallel$. We adopt the value of $r_{\rm min}=80~\rm pkpc$, which corresponds to the virial radius of dark matter haloes of mass $M_h\approx9\times10^{11}\rm~M_\odot$ \citep{2010ApJ...717..289S}. In Appendix~\ref{app:innermost}, we have tested the dependence of the Ly$\alpha$ transmission against the choice of the innermost radius from $r_{\rm min}=30\rm~pkpc$ to $160\rm~pkpc$ bracketing the value ($r_{\rm min}=1.5r_{\rm vir}$) suggested by \citet{2011ApJ...728...52L}.

The higher-order Lyman series scattering by the intervening circum-/inter-galactic gas also contributes to the opacity against the UV radiation from galaxies at the rest-frame wavelength $912~\mbox{\AA}<\lambda<1216~\mbox{\AA}$. The total effective optical depth by the line blanketing is $e^{-\tau_{\rm eff}^{\rm line}(\nu_e)}=\sum_{n=\alpha,\beta,\dots}e^{-\tau_{\rm eff}^{{\rm Ly}n}(\nu_e)}$. The calculation of the effective optical depth against the higher-order Ly$n$ line, $\tau_{\rm eff}^{{\rm Ly}n}(\nu_e)$, is identical to that of Ly$\alpha$ except for replacing the Ly$\alpha$ cross section with the higher-order Ly$n$ cross section.

\subsubsection{Ly$\alpha$ escape fraction}

We define the average total Ly$\alpha$ escape fraction, $\langle f_{\rm esc}^{\rm Ly\alpha}\rangle$, as a fraction of Ly$\alpha$ photons escaped from the sites of nebular Ly$\alpha$ emission within the ISM, through the CGM and IGM, and reached to a finite aperture of a telescope, averaged over many galaxies. Thus, the average total Ly$\alpha$ escape fraction is expressed as a product of the ISM Ly$\alpha$ escape fraction\footnote{The ISM Ly$\alpha$ escape fraction is defined as a fraction of Ly$\alpha$ photons escaped from the sites of nebular Ly$\alpha$ emission within the ISM to outside of galaxies.} and the Ly$\alpha$ transmissivity of the CGM and IGM, 
\begin{equation}
\langle f_{\rm esc}^{\rm Ly\alpha}\rangle=\langle f_{\rm esc,ISM}^{\rm Ly\alpha}\rangle\langle\mathcal{T}_\alpha\rangle.
\end{equation}

\subsection{Ly$\alpha$ haloes around galaxies}\label{sec:lya_halo}

While the various formation mechanisms of diffuse Ly$\alpha$ haloes are proposed, the physical origin of Ly$\alpha$ haloes is still unknown. The sources of Ly$\alpha$ emission in the diffuse haloes can be generated by (1) scattered Ly$\alpha$ photons from the central galaxies, (2) fluorescent Ly$\alpha$ radiation by the ionizing photons from the galaxies, (3) cooling radiation, and (4) Ly$\alpha$ emission from the unresolved galaxies around the detected galaxies. In this paper, we focus on the Ly$\alpha$ scattering contribution to Ly$\alpha$ haloes. 

Our perturbative approach can however be extended to include the contribution from the other Ly$\alpha$ emission such as fluorescence. We refer the reader to \citet{2017ApJ...841...19M} for discussion on the observational test for distinguishing different origins of Ly$\alpha$ haloes.

\subsubsection{$\LyA$ emissivity from scattered radiation}\label{sec:scattered_radiation}

In the first-order scattering expansion of cosmological Ly$\alpha$ transfer, we take into account the contribution from the photons that are scattered once. When Ly$\alpha$ haloes are formed due to the scattering of Ly$\alpha$ photons from the central sources, all the photons experienced no scattering (zeroth order expansion) do not contribute to the Ly$\alpha$ haloes. This is equivalent to the the {\it single scattering approximation}\footnote{Dijkstra \& Kramer (2012) has employed the single scattering approximation, but formulated differently from this paper. They have compared the analytic Ly$\alpha$ RT model based on the single scattering approximation with the Monte Carlo Ly$\alpha$ RT simulations, and showed that the single scattering approximation works well.}, in which we truncate the perturbative expansion in the first order. 

As Ly$\alpha$ photons are scattered by the neutral gas around galaxies, each absorber absorbs and re-emits the Ly$\alpha$ photons from the central $\LyA$ emitting galaxy with specific luminosity $L_\nu^{\rm intr}(\nu_e)=\langle L_\alpha^{\rm intr}\rangle\langle\Phi_\alpha^{\rm ISM}(\nu_e)\rangle$. Therefore, the surface brightness of Ly$\alpha$ haloes is given by integrating the Ly$\alpha$ emission from all the absorbers over the distribution and kinematics. 

First, we derive the individual absorber's luminosity illuminated by the scattering of $\LyA$ photons emitted from a central $\LyA$ emitting galaxy. The probability that an absorber absorbs $\LyA$ radiation is $1-e^{-\tau_{\rm a}}$. Only the area extended by an absorber can receive the radiation emitted from the central galaxy, therefore the Ly$\alpha$ radiation received by an absorber is $\sigma_{\rm abs}L_\nu^{\rm intr}/(4\pi l_p^2)$ where $\sigma_{\rm abs}$ is the geometrical cross section and $l_p=r/(1+z)$ is the proper distance. The absorber absorbs the emitted photons of frequency $\nu_e$ from the central galaxy by the amount of  $(1-e^{-\tau_{\rm a}})\sigma_{\rm abs}L_\nu^{\rm intr}(\nu_e)/(4\pi l_p^2 h\nu_\alpha)~[\rm photons~s^{-1}]$ and re-emits the photons near the Ly$\alpha$ line centre. 
 Thus, the individual absorber's Ly$\alpha$ luminosity (in units of $\rm erg~s^{-1}$) due to the scattering is given by,
\begin{equation}
L_\alpha^{\rm abs}(r,v_r,\NHI)= \int\left[1-e^{-\tau_{\rm a}(\nu_{\rm inj},\NHI)}\right]\frac{\sigma_{\rm abs}L_\nu^{\rm intr}(\nu_e)}{4\pi l_p(r)^2}d\nu_e,
\label{eq:absorber_luminosity}
\end{equation}
where $\nu_{\rm inj}=\nu_e\left[1-(H(z)l_p(r)+v_r\right)/c]$ is the frequency injected to an absorber.

Then, these individual absorber's Ly$\alpha$ luminosities integrate over all $\HI$ column density and velocities of the absorber. The velocity of an absorber affects the probability that Ly$\alpha$ photons from the central sources are scattered through its velocity-dependence in the optical depth. The bolometric Ly$\alpha$ emissivity $\langle\varepsilon_\alpha(r)\rangle$ (in units of $\rm erg~s^{-1}~cMpc^{-3}$) from the absorbers around galaxies is given by (see Appendix \ref{sec:A2} for derivation),
\begin{align}
&\langle\varepsilon_\alpha(r)\rangle=\frac{1+\xi(r)}{4\pi r^2}\int d\NHI\CDDF\left|\frac{dz}{dr}\right|\times \nonumber \\
&~~~~~~\int\frac{dv_r}{\sqrt{2\pi\sigma_v^2(r)}}\mathcal{L}_\alpha(r,v_r,\NHI)\exp\left[-\frac{(v_r-\langle v_r(r)\rangle)^2}{2\sigma_v^2(r)}\right]
\label{eq:Lya_emissivity}
\end{align}
where $\mathcal{L}_\alpha(r,v_r,\NHI)$ is the Ly$\alpha$ luminosity of the scattered photons by the absorbers around the central galaxies of specific luminosity $L_\nu$: 
\begin{equation}
\mathcal{L}_\alpha(r,v_r,\NHI)=\int \left[1-e^{-\tau_{\rm a}(\nu_{\rm inj},\NHI)}\right]L_\nu^{\rm intr}(\nu_e)d\nu_e.
\end{equation}

The calculation of the Ly$\alpha$ halo emissivity does not depend on the choice of the innermost radius $r_{\rm min}$. The Ly$\alpha$ halo profile at small radii, $r<r_{\rm min}$, can be self-consistently computed as long as we are concerned with the contribution from the singly scattered photons, which is a fundamental ansatz of the perturbative approach.

\subsubsection{The surface brightness of Ly$\alpha$ haloes}\label{sec:halo_model}

The mean surface brightness, $\langle {\rm SB}_\alpha(r_\perp)\rangle$, (in units of $\rm erg~s^{-1}~cm^{-2}~sr^{-1}$) of Ly$\alpha$ haloes due to the scattered Ly$\alpha$ photons from the central galaxies at impact parameter $r_\perp$ is then given by, 
\begin{equation}
\langle{\rm SB}_\alpha(r_\perp)\rangle=\frac{(1+z)^2}{4\pi(1+z)^4}\int_{r_\perp}^\infty\frac{2rdr}{\sqrt{r^2-r_\perp^2}}\langle\varepsilon_\alpha(r)\rangle.
\end{equation}
We apply the factor of $(1+z)^2$ to convert to the proper unit and the factor of $(1+z)^{-4}$ to convert it to the surface brightness value observed at $z=0$.

We also introduce a Ly$\alpha$ halo flux fraction $X_{\rm Ly\alpha,halo}$ as defined by \citet{2016A&A...587A..98W}, which is a fraction of the integrated Ly$\alpha$ flux in a Ly$\alpha$ halo relative to the sum of the central galaxy's Ly$\alpha$ line flux $F_{\rm gal}$ and the Ly$\alpha$ halo's flux $F_{\rm halo}$,
\begin{equation}
X_{\rm Ly\alpha,halo}=\frac{F_{\rm halo}}{F_{\rm gal}+F_{\rm halo}},
\end{equation}  
where the integrated Ly$\alpha$ halo flux is given by
\begin{equation}
F_{\rm halo}=\int \langle{\rm SB}_\alpha(r_\perp)\rangle\frac{2\pi r_\perp dr_\perp}{D_A^2(z)(1+z)^2 },
\end{equation}
and the Ly$\alpha$ line flux  of the central galaxy  is 
\begin{equation}
F_{\rm gal}=\langle f_{\rm esc, ISM}^{\rm Ly\alpha}\rangle\langle \mathcal{T}_\alpha\rangle\frac{\langle L_\alpha^{\rm intr}\rangle}{4\pi D_L^2(z)}.
\end{equation} 
$D_A(z)$ and $D_L(z)$ are the angular diameter and luminosity distance. Note that in our model the Ly$\alpha$ halo flux fraction is independent of $\langle L_\alpha^{\rm intr}\rangle$ as the increase of the intrinsic Ly$\alpha$ luminosity increases both the Ly$\alpha$ halo surface brightness and the observed Ly$\alpha$ luminosity of galaxies by the same factor. Furthermore, our halo flux fraction model depends on the ISM Ly$\alpha$ escape fraction in a functional form $X_{\rm Ly\alpha,halo}=1/(1+C\langle f_{\rm esc, ISM}^{\rm Ly\alpha}\rangle)$ where $C\propto \langle \mathcal{T}_\alpha\rangle/\int \langle{\rm SB}_\alpha(r_\perp)\rangle  2\pi r_\perp dr_\perp$ is the CGM/IGM-dependent number. The $\langle f_{\rm esc, ISM}^{\rm Ly\alpha}\rangle$ parameter only enters in the observed galaxy flux because the $\HI$ gas in the dust-free ISM only scatters the Ly$\alpha$ photons out of an observer's line-of-sight (but does not permanently destroy the photons), thus the scattered Ly$\alpha$ photons eventually find the way into the CGM and power Ly$\alpha$ haloes.

Overall, the above perturbative approach self-consistently models both Ly$\alpha$ in emission and absorption. The CGM/IGM around galaxies are traced by the cross-correlation between Ly$\alpha$ forests and galaxies. Thus, armed with this cosmological Ly$\alpha$ transfer calculation, we can study how the CGM and IGM affect the Ly$\alpha$ emission line profiles and Ly$\alpha$ haloes as a function of the observationally-calibrated gaseous environments.

\section{Physical Model of the gaseous environments around galaxies}\label{sec:model}

Finally, to realise the constrained RT technique we need to specify a parameterised model of the CGM and IGM around galaxies, which will be fitted to Ly$\alpha$ absorption observations. In this section, we describe a phenomenological model of the CGM and IGM. More specifically, our perturbative approach to Ly$\alpha$ RT requires models of ({\it i}) the $\HI$ column density distribution function, $\CDDF$, ({\it ii}) the real-space galaxy-absorber correlation function, $\xi(r)$, ({\it iii}) the average velocity field, $\langle v_r(r)\rangle$, and velocity dispersion, $\sigma_v(r)$, between galaxies and absorbers. These models are independent to the perturbative expansion of the cosmological Ly$\alpha$ transfer introduced in \S \ref{sec:theory}. Any physical model can be used. To illustrate the joint Ly$\alpha$ emission-absorption analysis we adopt a simple model in this paper. 

The structure of the CGM/IGM around galaxies is determined by both the structure of the intergalactic filaments in the large-scale structure and the hydrodynamical or radiative interaction of the gas with galaxies. Thus, we divide the modelling of the real-space clustering of the CGM/IGM around galaxies and its kinematics into two contributions: (1) from cosmological structure formation and (2) from the photo-ionization feedback and galactic winds from galaxies.

\subsection{The spatial distribution of gas around galaxies}\label{sec:real-space_clustering}

In our phenomenological model, the real-space correlation function between $\HI$ gas and galaxies is divided into the two contributions: (1) from cosmological structure formation (\S \ref{sec:grav}) and (2) from the photo-ionization feedback (\S \ref{sec:photo}). The resulting correlation function is expressed as,
\begin{equation}
1+\xi(r)=C^{\rm phot}(r)\left[1+\xi^{\rm grav}(r)\right],
\end{equation}
where $\xi^{\rm grav}(r)$ is the galaxy-absorber correlation function resulting from the gravitational clustering through the structure formation, whereas $C^{\rm phot}(r)$ is the correction factor for the clustering due to the photoionization feedback by galaxies.

\subsubsection{Gravitational clustering}\label{sec:grav}

The gravitational clustering of gas around galaxies is modelled using the power-law form with a slope $\gamma$ and a correlation length $r_c(z)$, 
\begin{equation}
\xi^{\rm grav}(r,z)=\left[\frac{r}{r_c(z)}\right]^{-\gamma}.
\end{equation}
Note that we have adopted the self-similar clustering ansatz, where the time-dependence of the gas-galaxy clustering only enters as a rescaling of the correlation length as a function of redshift, $r_c(z)$, while the functional shape is independent of redshift. Although the assumption of the self-similar clustering is only a crude approximation, it permits us to analytically capture the features of the full nonlinear growth of the gravitational clustering \citep{1985ApJS...58...39B,2003MNRAS.341.1311S}.

\subsubsection{Photoionization and UV background}\label{sec:photo}
The photoionization of the surrounding gas by the LyC photons ($\lambda<912$~\AA) from the galaxies lowers the number density of the absorbers around galaxies. As shown in Appendix~\ref{app:CDDF}, the impact of local ionizing sources introduces the correction factor,
\begin{equation}
C^{\rm phot}(r)=\left[1+\left(\frac{r}{r_{\rm ph}}\right)^{-2}\right]^{-\beta_{\rm eff}+1},\label{eq:correction_factor1}
\end{equation}
where $\beta_{\rm eff}$ is the effective slope of the CDDF when fit by a single power-law.  The equality radius $r_{\rm ph}$ is the radius at which the the local contribution, $\Gamma_{\rm local}(r)$, to photoionization rate is equal to the average photoionization rate of the UV background. $\Gamma_{\rm bkg}$ is the average UV background, for which we assume the value of $\Gamma_{\rm bkg}=1.0\times10^{-12}\rm~s^{-1}$ consistent with the observed value of $0.86^{+0.30}_{-0.22}\times10^{-12}\rm~s^{-1}$ at $z=3$ \citep{2013MNRAS.436.1023B}. We use the CDDF function fit by \citet{2012ApJ...746..125H} at $z=3$ and the effective slope of $\beta_{\rm eff}=1.5$. Then the equality radius is given by
\begin{align}
r_{\rm ph}&=\sqrt{\frac{(1+z)^2}{4\pi\Gamma_{\rm bkg}}\int_{\nu_{912}}^\infty  \sigma_{\mbox{\tiny{HI}}}(\nu)\frac{L_\nu(\nu)}{h\nu}d\nu}\label{eq:correction_factor2}
,  \\
&\approx 133{h^{-1}\rm{ckpc}}\left(\frac{1+z}{4}\right)\left(\frac{\langle f_{\rm esc}^{\rm LyC}\rangle}{0.02}\right)^{\frac{1}{2}}\left(\frac{\langle\rm SFR\rangle}{34~\rm{M_\odot yr^{-1}} }\right)^{\frac{1}{2}},\nonumber
\end{align}
where $\sigma_{\mbox{\tiny{HI}}}(\nu)=\sigma_{912}(\nu/\nu_{912})^{-3}$, $\sigma_{912}=6.304\times10^{-18}~\rm cm^2$, and $\nu_{912}$ is the frequency at the Lyman limit. We assumed the spectral energy distribution of $L_\nu(\nu)=h\alpha\dot{N}_{\rm ion}(\nu/\nu_{912})^{-\alpha}$ with a power-law EUV slope of $\alpha=3$, and used equation (\ref{eq:dNiondt}) for the LyC photon production rate, $\dot{N}_{\rm ion}$. We use $r_{\rm ph}=133h^{-1}\rm ckpc$ (i.e. $\langle f_{\rm esc}^{\rm LyC}\rangle=0.02$ and $\langle\rm SFR\rangle=34~\rm{M_\odot yr^{-1}}$) for the fiducial model.

In addition to the local ionizing sources, the large-scale UV background fluctuation can also modulate the number density of absorbers around galaxies. Because the length scale of the UV background fluctuations is of order of the mean free path of ionizing photons, $\lambda_{912}\approx346[(1+z)/4]^{-4.4}h^{-1}\rm cMpc$ \citep{2014MNRAS.445.1745W} and is much larger than the length scale of the CGM and IGM around galaxies, we model the impact of the large-scale UV background fluctuations by rescaling the CDDF. Following \citet{1997ApJ...486..599H} (see Appendix \ref{app:CDDF}), the CDDF responds as a function of the photoionization rate of the UV background,
\begin{equation}
\CDDF=\left.\CDDF\right|_{\bar{\Gamma}_{\rm bkg}(z)}\left[\frac{\Gamma_{\rm bkg}}{\bar{\Gamma}_{\rm bkg}(z)}\right]^{-\beta_{\rm eff}+1}.
\end{equation}
$\bar{\Gamma}_{\rm bkg}(z)$ is the average UV background of the entire Universe, for which we assume the value of $\bar{\Gamma}_{\rm bkg}(z=3)=1.0\times10^{-12}\rm~s^{-1}$.

\subsection{The kinematics of absorbers around galaxies}\label{sec:kinematics_model}

The cosmological inflow of the gas onto galaxies is predominantly controlled by the gravitational interaction through the large-scale structure formation. We can then write a Boltzmann equation for the pairs of galaxies and absorbers (Lagrangian gas parcels). By applying the method of BBGKY hierarchy \citep{1977ApJS...34..425D}, without loss of generality, the resulting first moment of the BBGKY hierarchy describes the conservation law of the galaxy-absorber pairs. This implies that the cosmological inflow is described by the average velocity field of absorbers around galaxies as
\begin{equation}
\langle v_r(r)\rangle=\frac{H(z)}{1+\xi^{\rm grav}(r,z)}\frac{\partial}{\partial z}\int_0^r\xi^{\rm grav}(r',z)\left(\frac{r'}{r}\right)^2dr'.\label{eq:BBGKY}
\end{equation}
While the BBGKY hierarchy is exact, in order to provide an analytic function for the average velocity field, we further assume the self-similar clustering ansatz (\S \ref{sec:real-space_clustering}). Solving equation (\ref{eq:BBGKY}), we find
\begin{equation}
\langle v_r(r)\rangle=-v_{\rm inflow}\frac{2(r/r_c)}{1+(r/r_c)^\gamma}.
\end{equation}
We introduced a free parameter $v_{\rm inflow}$ to parameterise the maximum average inflow velocity of gas onto galaxies.\footnote{In the self-similar clustering model, the solution to the BBGKY hierarchy follows $v_{\rm inflow}=\frac{1}{2}\frac{\gamma}{3-\gamma}\frac{d\ln r_c}{d\ln a}\frac{Hr_c}{1+z}$, which depends on the time-dependence of the correlation length, $\frac{d\ln r_c}{d\ln a}$. We expect the value of $d\ln r_c/d\ln a$ is order unity $\sim\mathcal{O}(1)$. For example, in the linear perturbation theory, the correlation length is linearly proportional to the scale length $d\ln r_c/d\ln a=1$. Therefore, the inflow velocity parameter is $v_{\rm inflow}\approx63{\rm~km~s^{-1}}(r_c/1 h^{-1}\rm cMpc)$ at $z=3$ and $\gamma=1.85$.} For the velocity dispersion of gas around galaxies, we assume the constant velocity dispersion $\sigma_v(r)=\rm const$ for simplicity.

The galactic outflow also affect the gas kinematics around galaxies, especially at inner radii close to the central galaxies. We have tested our model with outflow by launching the wind at a constant velocity at $\sim200\rm~km~s^{-1}$, which is consistent with the value used for Ly$\alpha$ ISM model of galaxies (\S \ref{sec:small}) and the median outflow velocity measured from the interstellar metal absorption lines \citep{2010ApJ...717..289S}. The region of mechanical influence is out to radius $\approx200\rm~pkpc$ which is the maximum radius that gas can be inertially transported during $1\rm~Gyr$. The result was only marginally affected; thus, we avoid introducing an extra parameter for outflow in the CGM and IGM in this paper.

\section{Joint Analysis of galaxy-Ly$\alpha$ absorption clustering with Ly$\alpha$ emission}\label{sec:analysis}

\subsection{Ly$\alpha$ absorption around galaxies}\label{sec:calibration}

\begin{figure}
\centering
  \includegraphics[angle=0,width=0.85\columnwidth]{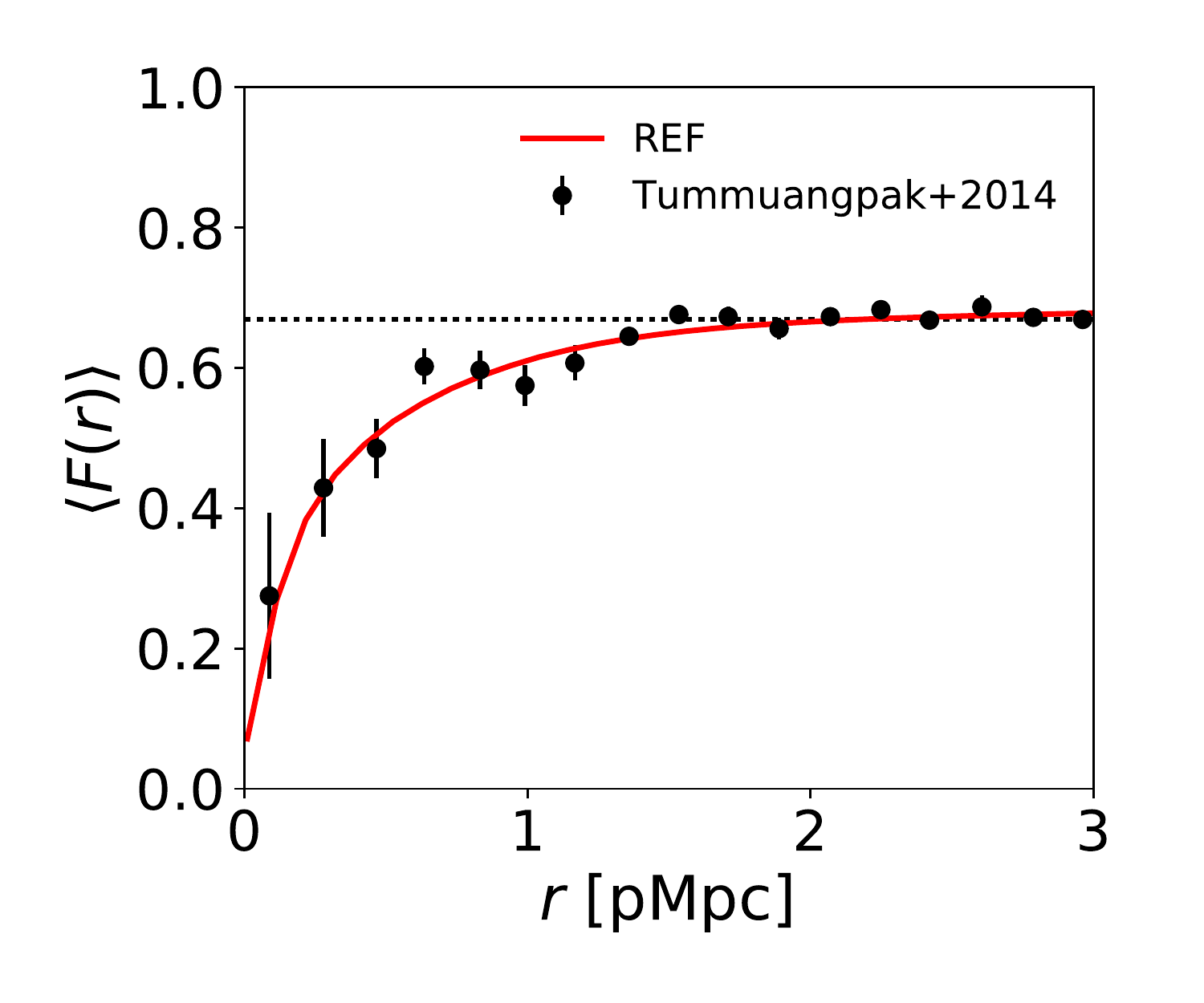}
  \includegraphics[angle=0,width=0.8\columnwidth]{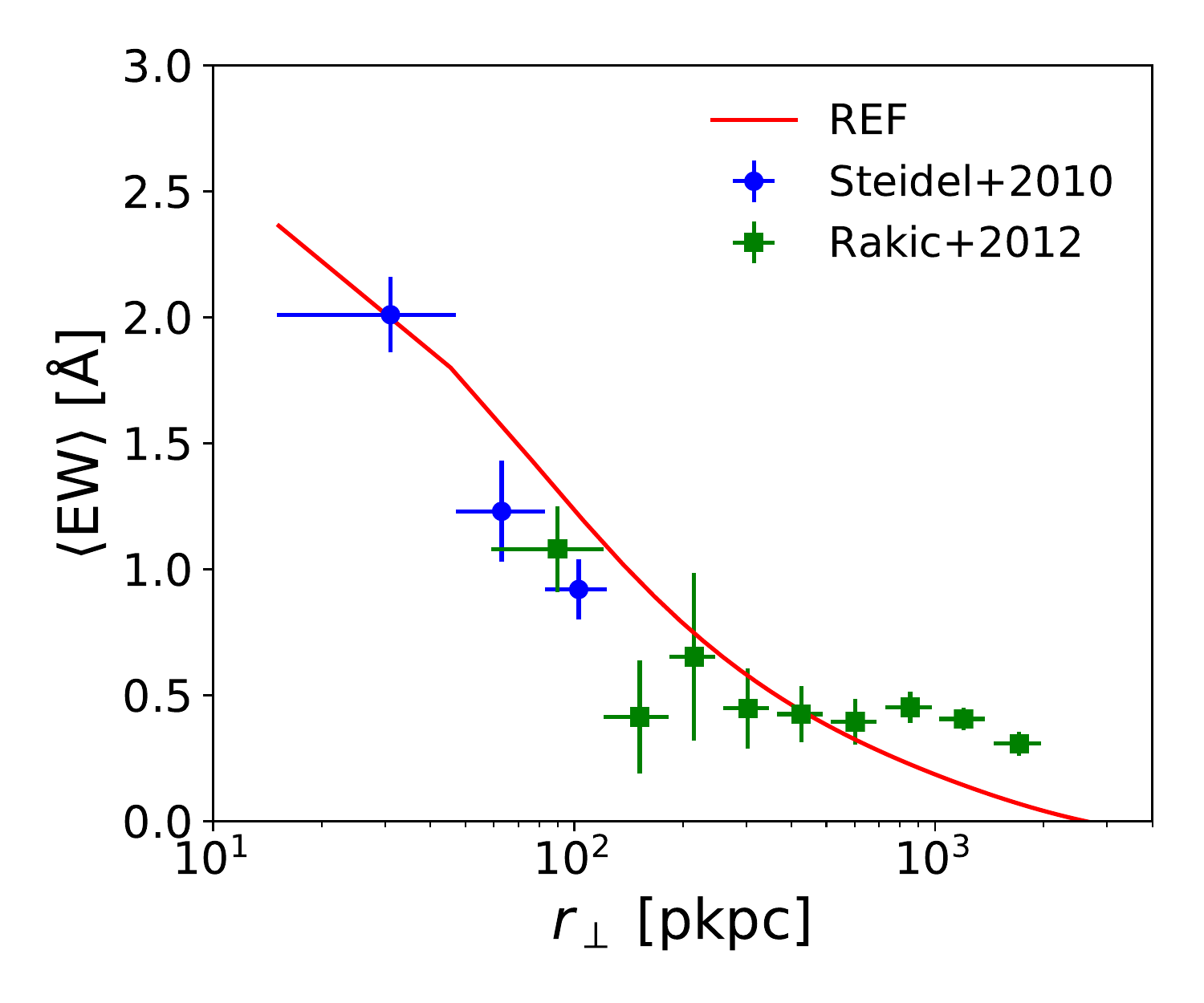}
  \caption{Comparison of the REF model with data. {\bf Top panel}: The spherically averaged transmitted Ly$\alpha$ transmitted flux of the REF model (red solid) as a function of radius. The black data points are the measurement from VLRS survey (Tummugapak et al. 2014). The dotted line is the mean Ly$\alpha$ transmitted flux of the IGM (Becker et al. 2013).  {\bf Bottom panel}: The excess equivalent width of Ly$\alpha$ absorption as a function of impact parameter $r_\perp$ for the REF model (red solid). The blue and green data points are the measurements from galaxy-galaxy pairs (Steidel et al. 2010) and galaxy-QSO pairs (Rakic et al. 2012) in KBSS survey. }
   \label{fig:calibration}
\end{figure}

\subsubsection{Parameter calibration and observational test}

We calibrate our model parameters of the CGM and IGM around galaxies introduced in \S \ref{sec:model} using the observational data from the spectroscopic galaxy surveys in the QSO fields: VLT LBG Redshift Survey \citep{2011MNRAS.414...28C,2014MNRAS.442.2094T,2017MNRAS.471.2174B}, \citet{2003ApJ...584...45A,2005ApJ...629..636A}, and Keck Baryonic Structure Survey \citep{2010ApJ...717..289S,2012ApJ...751...94R,2014MNRAS.445..794T}. We choose the best-fit correlation length $r_c$ and slope $\gamma$, the inflow velocity parameter $v_{\rm inflow}$, and the velocity dispersion parameter $\sigma_v$, by comparing the model prediction of the 2D galaxy-Ly$\alpha$ forest cross-correlation function with observed one. We also use the equivalent width of Ly$\alpha$ absorption around galaxies by \citet{2010ApJ...717..289S,2012ApJ...751...94R} to complement the calibration procedure. We then find the best-fit parameters by fitting the model by eye. 

The CDDF, $\CDDF$, and photoionization rate of the UV background, $\Gamma_{\rm bkg}$, are pre-determined based on the observation and analysis of QSO absorption spectra (see \S \ref{sec:real-space_clustering}).

To find the best-fit parameters of $r_c$ and $\gamma$, we first compare the model with the spherically averaged Ly$\alpha$ transmitted flux and the equivalent width of Ly$\alpha$ absorption. Figure \ref{fig:calibration} shows the result of the calibration. The resulting parameters are listed in Table~\ref{table:model} and we refer to this set of parameters as REF model in the rest of the paper.

\begin{table}
\centering
\caption{The model parameters of star-forming galaxies and the CGM/IGM.}
\label{table:model}
\begin{tabular}{ll}
\hline\hline
\multicolumn{2}{c}{--- ISM parameters ---} \\
 Star formation rate & $\langle\rm SFR\rangle=34~\rm M_\odot yr^{-1}$ \\
 Average ISM Ly$\alpha$ escape fraction & $\langle f_{\rm esc, ISM}^{\rm Ly\alpha}\rangle=0.20$ \\
 Average LyC escape fraction & $\langle f_{\rm esc}^{\rm LyC}\rangle=0.02$ \\
 ISM Ly$\alpha$ line profile & Gronke \& Dijkstra (2016) model \\
 \\
\multicolumn{2}{c}{--- CGM/IGM parameters ---} \\
 Correlation length & $r_c=1.0h^{-1}\rm cMpc$ \\
 Power-law slope & $\gamma=1.85$ \\
 Photoionization rate of UV background & $\Gamma_{\rm bkg}=1\times10^{-12}\rm~s^{-1}$ \\
 Inflow velocity & $v_{\rm inflow}=135\rm~km~s^{-1}$ \\
 Velocity dispersion & $\sigma_v=200\rm~km~s^{-1}$ \\
\hline\hline
\end{tabular}
\end{table} 

\begin{figure}
\advance\leftskip-0.2cm
\includegraphics[angle=0,width=1.05\columnwidth]{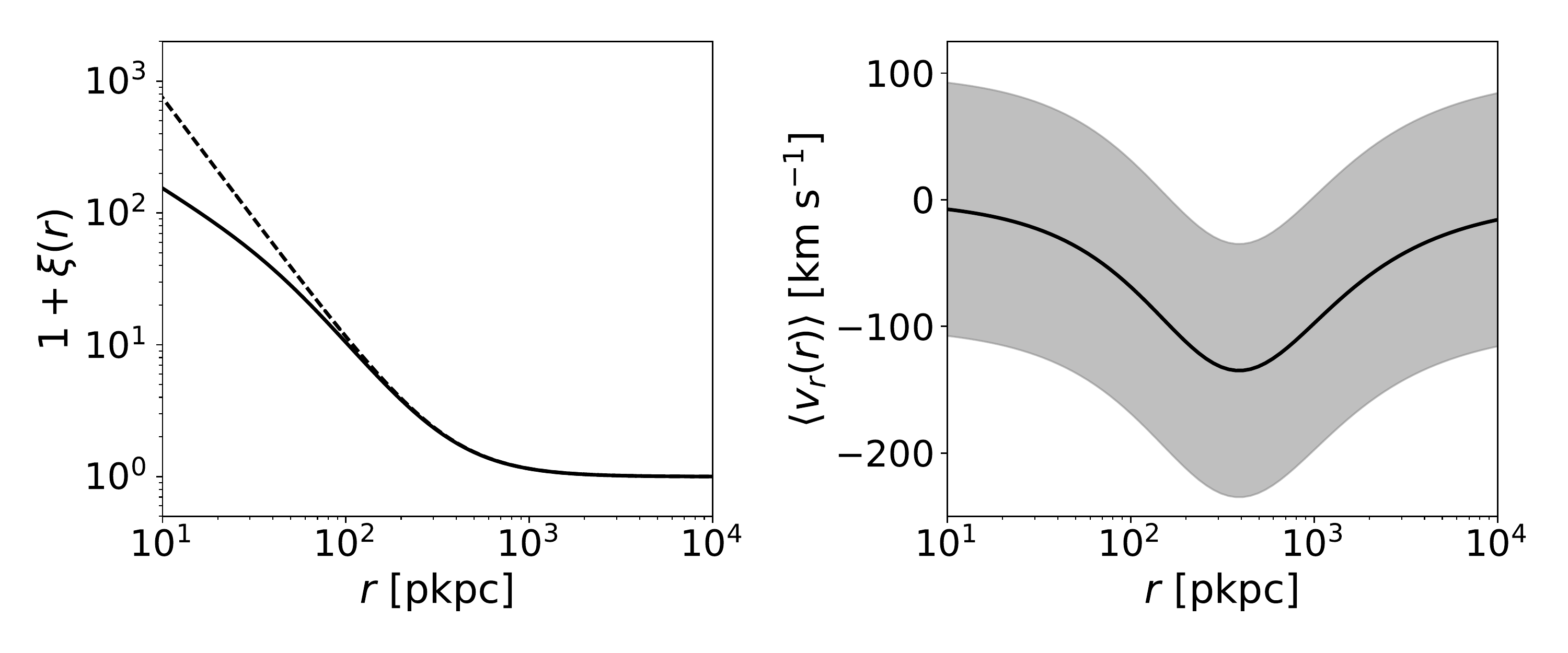}
\vspace{-0.3cm}
\caption{The reference model of the CGM and IGM around galaxies. {\bf Left panel:} The real-space correlation function of $\HI$ gas around galaxies of the REF model with the photoionization feedback (solid). The dotted line shows the case for the gravitational clustering without photionization feedback. {\bf Right panel:} The average peculiar velocity field of gas around galaxies. The solid line is the cosmological inflow based on self-similar clustering ansatz. The width of the grey shaded region indicates $1\sigma$ scatter due to the velocity dispersion in the Gaussian streaming model.}\label{fig:model}
\end{figure}

Figure~\ref{fig:model} (left panel) plots the best-fit real-space correlation function between absorbers and galaxies (solid curve), which include both contributions from gravitational clustering and photoionization feedback. As a reference, the dashed curve shows the correlation function without photoionization feedback. The clustering of the gas around galaxies increases the abundance of the absorbers, which extends out to $\sim1\rm~pMpc$. Compared with the purely gravitational contribution, the photoionization feedback by the central galaxies mildly lowers the abundance of the neutral gas within $\lesssim100\rm~pkpc$.

The gas kinematics is poorly constrained by the available Ly$\alpha$ absorption data. While VLRS reports the measurement of the 2D galaxy-Ly$\alpha$ forest cross-correlation at $z=3$ \citep{2014MNRAS.442.2094T,2017MNRAS.471.2174B}, the tabulated data to be directly compared with the model was not available at the time of writing. \citet{2012ApJ...751...94R,2014MNRAS.445..794T} present the tabulated data, but in terms of the median optical depth. As we cannot directly compare our model with these measurements,
we choose a inflow velocity parameter, $v_{\rm inflow}$, by comparing with cosmological hydrodynamic simulations of \citet{2014MNRAS.445.2462M,2015MNRAS.453..899M} that are tested against the observed Ly$\alpha$ absorption signal around galaxies \citep{2012ApJ...751...94R}. \citet{2014MNRAS.445.2462M} reports the mean radial velocity profile of gas around haloes with total mass of $4.5\times10^{11}\rm~M_\odot$ at $z=3$. The simulations show the gas inflow of velocity $\sim100-150\rm~km~s^{-1}$ around $\sim1-10\rm~cMpc$. We therefore choose the inflow velocity parameter of $v_{\rm inflow}=135~\rm km~s^{-1}$ for our REF model.\footnote{Near the completing of our work, \citet{2017MNRAS.471..690T} independently reported a comparison of the observed 2D median optical depth map with the EAGLE simulations, and find a similar infall velocity (velocity dispersion) to our phenomenological model. This assures the fidelity of our model and calibration.}

\begin{figure}
 \begin{center}
  \includegraphics[angle=0,width=\columnwidth]{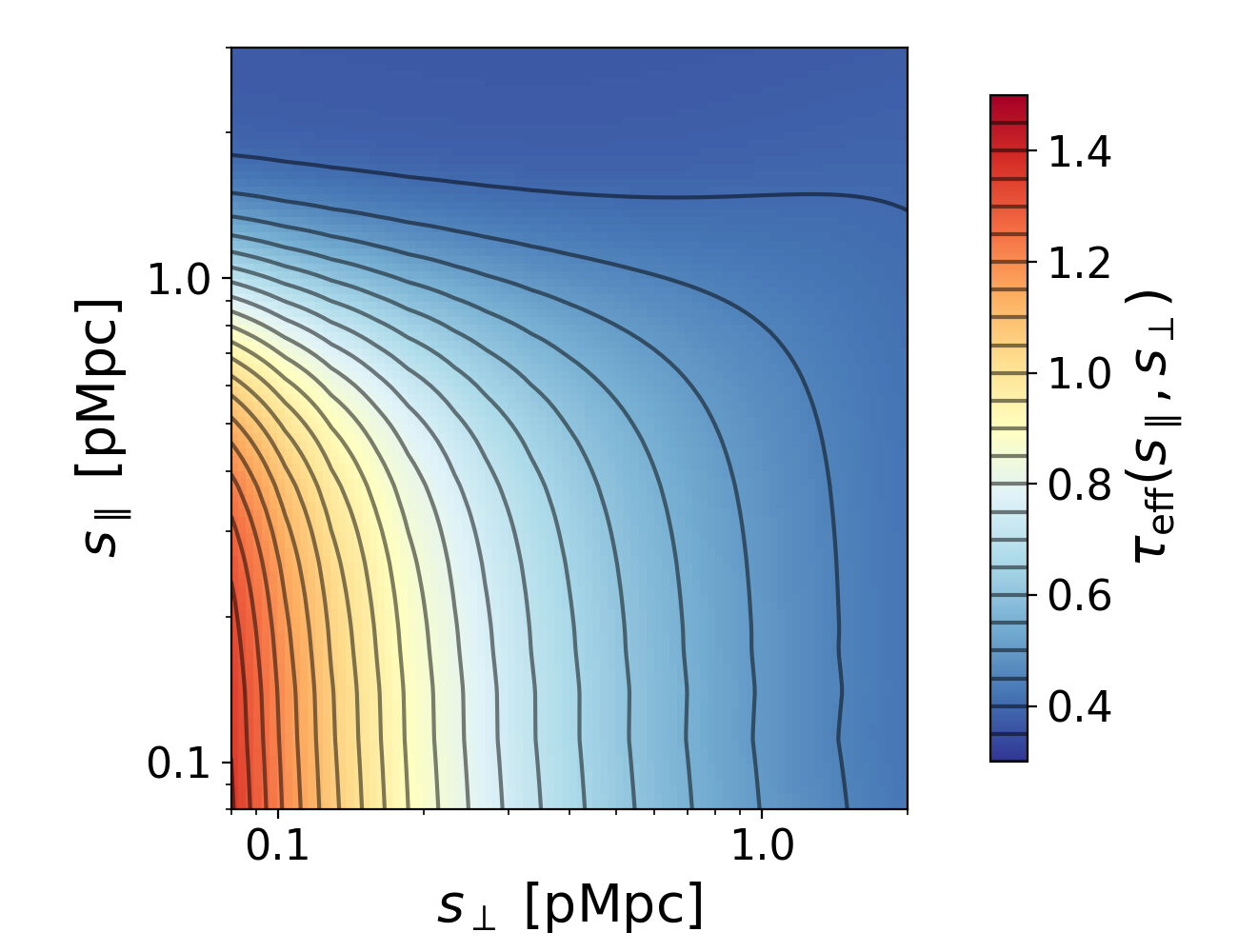}
  \caption{2D Ly$\alpha$ effective optical depth map around galaxies at $z=3$ as a function of the line-of-sight, $s_{\parallel}$, and the transverse, $s_{\perp}$, separations between the Ly$\alpha$ absorption pixels and galaxies in the REF model.}
   \label{fig:RSD_REF}
 \end{center}
\end{figure}

\begin{figure*}
 \begin{center}
\includegraphics[angle=0,width=\textwidth]{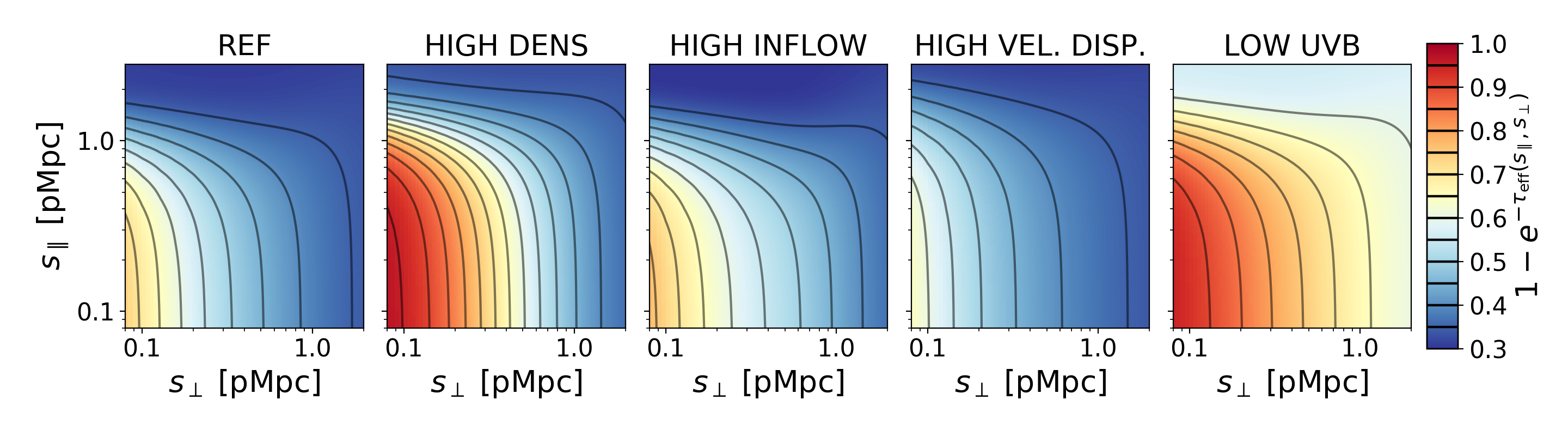}
   \label{fig:RSD_DENS}
   \vspace{-0.5cm}
     \caption{2D map of Ly$\alpha$ transmitted flux around galaxies. All the outermost contour is $1-e^{-\tau_{\rm eff}}=0.35$ and increases by $0.05$ inwards, except for LOW UVB model for which the outermost contour is $0.60$. The model parameters for REF model is shown in Table~\ref{table:model}. HIGH DENS model has $r_c=2h^{-1}\rm cMpc$, HIGH INFLOW model has $v_{\rm inflow}=300\rm~km~s^{-1}$, HIGH VEL. DISP. model has  $\sigma_v=300\rm~km~s^{-1}$, and LOW UVB model has $\Gamma_{\rm bkg}=2\times10^{-13}\rm~s^{-1}$. All other parameters are the same as REF model.}\label{fig:RSDs}
 \end{center}
\end{figure*}

Furthermore, we choose the constant velocity dispersion parameter of $\sigma_v=200\rm~km~s^{-1}$. This is based on the reported value of $240\pm60\rm~km~s^{-1}$ by \citet{2017MNRAS.471.2174B} derived by their model fitting procedure to the 2D galaxy-Ly$\alpha$ forest cross-correlation. Once the tabulated data of the 2D galaxy-Ly$\alpha$ forest cross-correlation is available, we can readily calibrate our model of the gas kinematics in a more consistent manner in future.

Figure~\ref{fig:model} (right panel) shows the average peculiar velocity field of gas around galaxies (solid curve) in our REF model. The gray shaded region indicates the velocity dispersion of the gas. The gas infalls from large-scale to small-scale by cosmological inflow as a result of structure formation. The mean peculiar gas velocity then approaches zero at inner radii inside the turnaround radius of the haloes. This trend is in agreement with the simulations of \citet{2012MNRAS.423.2991V,2014MNRAS.445.2462M}, where such deceleration is caused by the shocks across the multi-streaming gas and the virialization of the haloes. 

After the calibration, our model of the gas distribution and kinematics of the CGM and IGM around galaxies are constrained to be consistent with the galaxy-Ly$\alpha$ forest clustering data. Thus, the Ly$\alpha$ radiative transfer calculation using this calibrated model allows us to self-consistently study the interaction between Ly$\alpha$ emission line profiles and haloes of galaxies and the large-scale gaseous environment.

\subsubsection{2D redshift-space galaxy-Ly$\alpha$ forest clustering: \newline the imprints of the CGM and IGM}

Before proceeding to the Ly$\alpha$ emission properties, we present the 2D redshift-space galaxy-Ly$\alpha$ forest clustering, which contains the full statistical information of the clustering and kinematics of the CGM and IGM around galaxies. Figure~\ref{fig:RSD_REF} shows the 2D effective optical depth map $\tau_{\rm eff}(s_\parallel,s_\perp)=-\ln\langle F(s_\parallel,s_\perp)\rangle$ in the REF model calibrated in the previous section. The model clearly shows the excess of Ly$\alpha$ absorption near the central galaxies and the redshift-space anisotropy in the 2D effective optical depth map. The feature of the redshift-space anisotropy in the model strikingly resembles the one observed by \citet{2014MNRAS.445..794T}.

\begin{table}
\centering
\caption{A grid of the CGM/IGM models. The parameter only differs from the REF model is shown.}
\label{table:modelgrid}
\begin{tabular}{ll}
\hline
HIGH DENS & $r_c=2.0h^{-1}\rm cMpc$ \\
HIGH INFLOW & $v_{\rm inflow}=300\rm~km~s^{-1}$ \\
HIGH VEL. DISP.  & $\sigma_v=300\rm~km~s^{-1}$ \\
LOW UVB & $\Gamma_{\rm bkg}=2\times10^{-13}\rm~s^{-1}$ \\
\hline
\end{tabular}
\end{table}

With Figure~\ref{fig:RSDs} we show how the gas distribution and kinematics of the CGM and IGM around galaxies are traced by the 2D redshift-space galaxy-Ly$\alpha$ forest clustering. The variation of the 2D effective optical depth maps for the four different gaseous environments are shown in Figure \ref{fig:RSDs}. Each model is perturbed relative to the REF model with respect to gas density, HIGH DENS ($r_c=2h^{-1}\rm cMpc$), inflow velocity, HIGH INFLOW ($v_{\rm inflow}=300\rm~km~s^{-1}$), and velocity dispersion, HIGH VEL. DISP. ($\sigma_v=300\rm~km~s^{-1}$), and UV background, LOW UVB ($\Gamma_{\rm bkg}=2\times10^{-13}\rm~s^{-1}$) (tabulated in Table~\ref{table:modelgrid}). All the other parameters are the same as the REF model.

\begin{figure}
\centering
\includegraphics[angle=0,width=0.8\columnwidth]{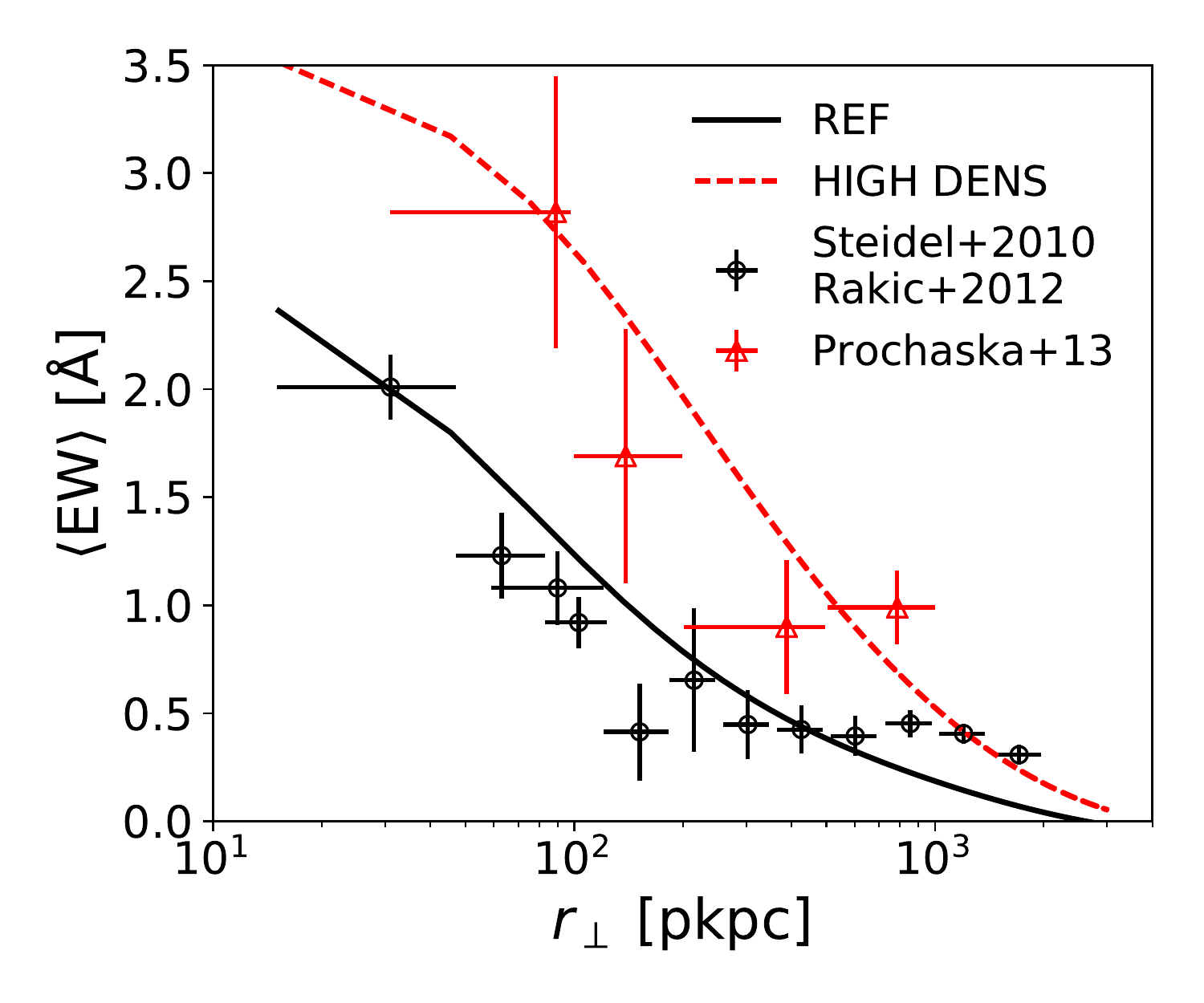}
\caption{Same as the bottom panel of Figure~\ref{fig:calibration}. The figure shows that the HIGH DENS (red dashed) model represents a high density environment similar to the Prochaska et al. (2013) observation (red triangles).}\label{fig:HIGH_DENS_EW}
\end{figure}

The $\HI$ gas density affects the amplitude of the 2D effective optical depth map. Our HIGH DENS model in Figure~\ref{fig:RSDs} shows that increasing the correlation length of gas around galaxies increases the effective optical depth near the centre. Such high density environment resembles the region around QSOs. This is illustrated in Figure~\ref{fig:HIGH_DENS_EW} where we compare the Ly$\alpha$ absorption equivalent width in our HIGH DENS model with the observation around QSOs \citep{2013ApJ...776..136P}. The photoionization similarly affects the amplitude (see LOW UVB model). While a higher LyC leakage from galaxies preferentially lowers the Ly$\alpha$ absorption closer to galaxies, a lower photoionization rate of the UV background uniformly increases the opacity of gas around galaxies, making the CGM and IGM more neutral.

On the other hand, the gas kinematics (inflow and velocity dispersion: HIGH INFLOW and HIGH VEL. DISP. models) impact the redshift-space anisotropy. On $\sim1\rm~pMpc$ scale, the clumpy neutral gas inflow onto galaxies causes the squashing of the effective optical depth map along the line-of-sight direction. Such large-scale inflow of the gas on to galaxies has been detected in observations \citep{2012ApJ...751...94R,2017MNRAS.471.2174B} in a consistent matter with cosmological simulations \citep{2017MNRAS.471..690T}. A higher velocity dispersion introduces a more line-of-sight elongation in the redshift-space anisotropy. Note that including outflow in the model will also introduce a similar elongation. However, in detail, the quantitative elongation signature by the increasing random gas motion differs from the coherent change in the velocity field by outflow. They should be distinguishable in principle.

\begin{figure*}
\centering
  \includegraphics[angle=0,width=\textwidth]{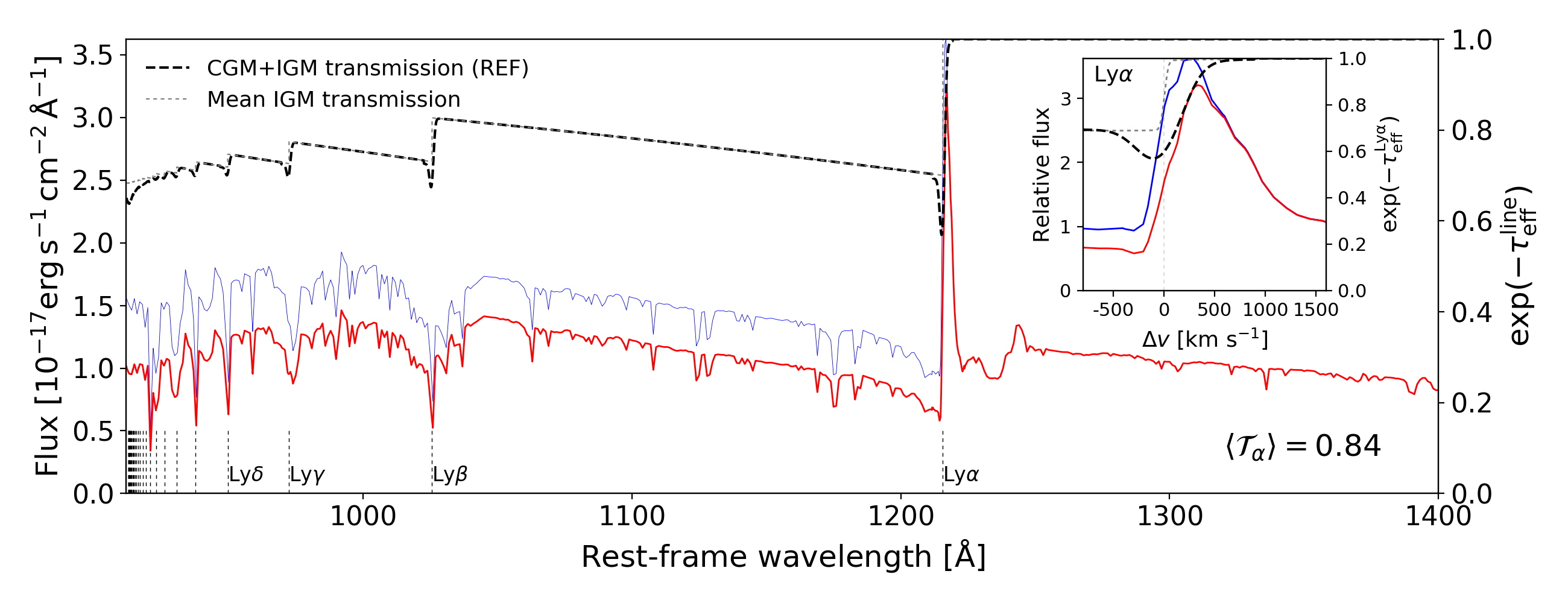}
   \vspace{-0.5cm}
  \caption{The composite spectrum (red solid, the flux in the left y-axis) of galaxies with the emergent Ly$\alpha$ line profile and the effective CGM/IGM transmission $e^{-\tau_{\rm eff}^{\rm line}}$ (black dashed, the  value is in the right y-axis) in the REF model. The blue line shows the spectrum without the effect of the CGM and IGM, using the BPASS population synthesis model \citep{2017arXiv171002154E} with constant star formation rate, $100\rm~Myr$ age, and metallicity $Z=0.001$ as an example. For a comparison, the effective IGM transmission without the CGM effect \citep{2006MNRAS.365..807M} is shown as the grey dotted line (mean IGM model). The value of Ly$\alpha$ transmission $\langle\mathcal{T}_\alpha\rangle$ is quoted at the bottom right corner. The inset shows the zoom-in plot of the Ly$\alpha$ line profile.}
   \label{fig:spectrum}
\end{figure*}

\subsection{Emergent Ly$\alpha$ emission line profile}
\subsubsection{The impact of the CGM and IGM predicted from the galaxy-Ly$\alpha$ absorption clustering data}

Once the model of the CGM and IGM is constrained by the galaxy-Ly$\alpha$ forest clustering measurements, the constrained Ly$\alpha$ RT approach self-consistently predicts the impact of the CGM/IGM on the emergent Ly$\alpha$ line profile of galaxies. 
Figure~\ref{fig:spectrum} shows the emergent average spectrum of star-forming galaxies and the Ly$\alpha$ line profile after propagating through the CGM and IGM in the REF model (red solid curve) [the intrinsic model galaxy spectrum (blue solid curve) is shown for a comparison]. The impact of the CGM/IGM is encapsulated in the CGM/IGM transmission curve (black dashed curve with right ordinate), which is a self-consistent estimate of $e^{-\tau_{\rm eff}^{\rm Ly\alpha}}$ (and $e^{-\tau_{\rm eff}^{\rm line}}$) (\S \ref{sec:large}) from the galaxy-Ly$\alpha$ forest clustering constraint without any adjustable free parameter.  

In the observationally constrained CGM/IGM, only $\langle\mathcal{T}_\alpha\rangle=0.84$ of the intrinsic Ly$\alpha$ flux is transmitted, while nearly $100$ per cent of photons can propagate toward observers in the absence of clustering (Mean IGM, grey dotted curve). This means that {\it the impact of the CGM/IGM on the Ly$\alpha$ line flux is small, but non-negligible even at $z\sim2-3$}. While the precise value of the transmission of course depends on the ISM Ly$\alpha$ line profile as $\langle\mathcal{T}_\alpha\rangle=\int e^{-\tau^{\rm Ly\alpha}_{\rm eff}}\langle \Phi_\alpha^{\rm ISM}\rangle d\nu_e$, this conclusion still holds. We will return to this point in \S~\ref{sec:escape}. The inset of Figure~\ref{fig:spectrum} shows in detail the CGM/IGM transmission curve near the Ly$\alpha$ line. The clustering of neutral gas around galaxies scatters more Ly$\alpha$ photons out of our line of sight, causing the ``attenuation dip" near the Ly$\alpha$ resonance line centre \citep{2011ApJ...728...52L}. The shape of this attenuation dip is determined by the $\HI$ gas distribution and kinematics of the CGM. We will examine this in detail in the following section.

Furthermore, the clustering of gas in the CGM imprints the excess attenuation near the higher-order Lyman series in the average galaxy spectrum, adding a series of attenuation dips by the CGM \citep{2013ApJ...769..146R} on top of the characteristic sawtooth-shaped IGM transmission curve \citep{1995ApJ...441...18M,2006MNRAS.365..807M,2014MNRAS.442.1805I}. The shape of the attenuation dips near the higher-order Lyman series is well correlated with that of Ly$\alpha$, which only differs by the difference in the oscillator strengths of the atomic transitions. Note that the intrinsic absorption at stellar atmosphere of massive stars (in the stellar population synthesis model) also causes the attenuation at the higher-order Lyman series lines. Thus, the scattering of the higher-order Lyman series photons by the CGM causes excess attenuations at the higher order Lyman series over the intrinsic stellar atmosphere and ISM features. The higher-order Lyman series features in the average spectrum may be used as a consistency check of the impact of the CGM and IGM on the emergent Ly$\alpha$ line.

 \subsubsection{Dissecting the impacts of the CGM and IGM: \newline the clustering and kinematics of the gas around galaxies}\label{sec:5.2.2}
 
\begin{figure*}
 \begin{center} 
    \includegraphics[angle=0,width=\textwidth]{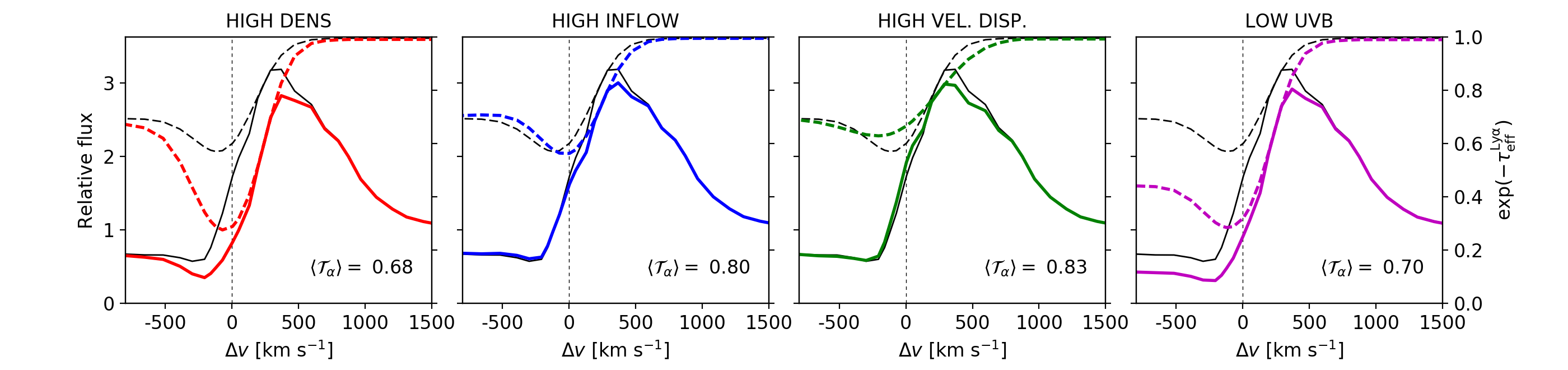}
    \vspace{-0.5cm}
  \caption{Comparison of the average emergent Ly$\alpha$ line profiles in HIGH DENS (red solid), HIGH INFLOW (blue solid), HIGH VEL. DISP. (green solid) and LOW UVB (purple solid) models with REF model (black solid). The value of the relative flux in each line profile is shown in the left y-axis. The CGM/IGM transmission curve for each model is indicated as dashed lines with a value shown in the right y-axis. The value of Ly$\alpha$ transmission in each model is shown at the bottom right corner. HIGH DENS model has $r_c=2h^{-1}\rm cMpc$, HIGH INFLOW model has $v_{\rm inflow}=300\rm~km~s^{-1}$, HIGH VEL. DISP. model has  $\sigma_v=300\rm~km~s^{-1}$, and LOW UVB model has $\Gamma_{\rm bkg}=2\times10^{-13}\rm~s^{-1}$. All other parameters are the same as REF model.}
   \label{fig:lineprofiles}
 \end{center}
\end{figure*}

We now examine the details of how the changes of gas density, and kinematics, and UV background in the CGM and IGM around galaxies traced by the galaxy-Ly$\alpha$ forest clustering (Figure~\ref{fig:RSDs}) impact the Ly$\alpha$ line profiles. Figure~\ref{fig:lineprofiles} shows the emergent Ly$\alpha$ line profiles after propagating through the high gas density (HIGH DENS), high inflow velocity (HIGH INFLOW), high velocity dispersion (HIGH VEL. DISP.), and low UV background (LOW UVB) model environments.

A stronger clustering of gas around galaxies enhances the attenuation of Ly$\alpha$ just blueward of the line centre, which is responsible for creating a dip of the attenuation in the transmission curve. Ly$\alpha$ line transfer is sensitive to total velocity field of the gas (i.e. Hubble flow + peculiar velocity). Including the Hubble flow, most of the gas is in fact experiencing net outflow. As the excess distribution of gas by the clustering extends out to $\sim1\rm~pMpc$ around galaxies where the Hubble flow is $\sim300\rm~km~s^{-1}$ at $z\sim3$, the dip ranges approximately over $\Delta v\approx0-300{\rm~km~s^{-1}}$. The attenuation dip increases with increasing gas density around galaxies. In a high density environment, the Ly$\alpha$ flux is suppressed even more. Only $\langle\mathcal{T}_\alpha\rangle=0.68$ of the Ly$\alpha$ flux is transmitted through the CGM and IGM in the high density environment, while $\approx0.84$ of Ly$\alpha$ flux can be transmitted through the REF model of the CGM and IGM around galaxies. 

The (wavelength-dependent) shape of the CGM/IGM transmission curve near the Ly$\alpha$ line centre is affected by the $\HI$ gas kinematics around galaxies. The inflowing gas causes the attenuation in the redward ($\Delta v>0$) of the line centre whereas the outflowing gas attenuates the blueward ($\Delta v<0$). As the gas infalls onto galaxies, the velocity field deviates from the Hubble flow, resulting the net inflow of the gas. This inflowing gas can scatter Ly$\alpha$ photons intrinsically emitted at the redward of the line from the central galaxies as the photon is seen blueshifted to the line centre in the rest-frame of the inflowing gas. Therefore, as the inflowing gas attenuates the Ly$\alpha$ photons, it produces an excess Ly$\alpha$ attenuation at the redward of the line centre. Thus, a higher gas inflow velocity suppresses more Ly$\alpha$ flux at the redward of the line centre. For example, increasing the inflow velocity parameter from $135\rm~km~s^{-1}$ to $v_{\rm inflow}=300\rm~km~s^{-1}$ (HIGH INFLOW model) increases the attenuation of the redshifted Ly$\alpha$ line to $\langle\mathcal{T}_\alpha\rangle\approx0.80$, making the CGM and IGM more susceptible of scattering the redshifted Ly$\alpha$ photons emitted from galaxies.
 
Furthermore, the velocity dispersion of the gas around galaxies impacts the width of the attenuation dip. Because of the velocity dispersion of gas around galaxies, there are both outflowing and infalling gas along with the coherent flow. Such fast moving outliers of the gas broaden the width of the attenuation dip. A higher gas velocity dispersion of the CGM and IGM lowers the attenuation close to the line centre. This is because the chaotic motion of the gas creates the path of escapes in the gas around galaxies, which is otherwise opaque to Ly$\alpha$ photons because of the coherent inflowing gas. On the other hand, at larger radii, where the gas is predominantly outflowing, a higher velocity dispersion means that some gas can be blueshifted into the line centre and scatter the Ly$\alpha$ photons, otherwise the coherent Hubble flow lets the photons to escape. This creates more attenuation at the redward of the line centre than the lower velocity dispersion case. 

Finally, a change in the UV background affects the overall Ly$\alpha$ transmission through the CGM and IGM around galaxies. A lower photo-ionization rate makes gas to be more neutral. This increase in the residual neutrality of the gas increases the abundance of the $\HI$ absorbers around galaxies. Because the effective optical depth scales as $\tau^{\rm Ly\alpha}_{\rm eff}\propto\Gamma_{\rm bkg}^{-1/2}$, a lower photo-ionization rate of the UV background attenuates more Ly$\alpha$ photons from galaxies. For example, lowering the photoionization rate from $\Gamma_{\rm bkg}=1\times10^{-12}\rm~s^{-1}$ to $\Gamma_{\rm bkg}=2\times10^{-13}\rm~s^{-1}$ decreases the Ly$\alpha$ transmission from $\langle\mathcal{T}_\alpha\rangle\approx0.84$ to $\langle\mathcal{T}_\alpha\rangle\approx 0.70$. 

\subsubsection{Relative contribution of absorbers}

Different types of absorbers, $\LyA$ forest absorbers with $\NHI<10^{17}\rm~cm^{-2}$ and Lyman-limit systems/damped Ly$\alpha$ absorbers (LSS/DLAs) with $\NHI>10^{17}\rm~cm^{-2}$, contribute differently to the CGM/IGM transmission curve. At the redward of $\LyA$ line centre, the formation of the red damping wing is mainly driven by the two contributions: (1) infalling Ly$\alpha$ forest absorbers onto galaxies and (2) the LLS/DLAs around galaxies. 

As discussed in \S~\ref{sec:5.2.2}, the infalling absorbers scatter Ly$\alpha$ photons emitted at the redward of line centre. The contribution to the transmission curve extends to the maximum velocity of infalling Ly$\alpha$ forest absorbers. This could be larger than the maximum mean inflow velocity parameter $v_{\rm inflow}$ due to the velocity dispersion. As a result of the chaotic gas kinematics (large $\sigma_v$) in the CGM, the CGM/IGM transmission curves extend more smoothly to larger positive $\Delta v$ than a case of the coherent gas kinematics with a lower velocity dispersion  (small $\sigma_v$) . In fact, at $z=3$ the contribution of infalling Ly$\alpha$ forest absorbers dominates the formation of the red damping wing opacity (see Appendix \ref{app:absorbers}).

While the Ly$\alpha$ forest absorbers contribute to the red damping wing when the gas is inflowing, the LLS/DLAs around galaxies can contribute the damping wing even when the gas is outflowing. Since the LLS/DLAs show the Lorentz wing absorption due to the strong Ly$\alpha$ absorption, the Lorentz wing scattering by the outflowing LLS/DLAs contributes to the redward of line centre. This effect is subdominant in the damping wing at $z=3$ because the number density of LLS/DLAs is low. However, when the photoionization rate of the UV background is lower, the increasing number density of LLS/DLAs causes a non-negligible contribution, and can dominate the formation of the red damping wing opacity.  The result is presented in Appendix \ref{app:absorbers}.

\subsection{Ly$\alpha$ escape fraction}\label{sec:escape}

\begin{figure}
  \centering
   \includegraphics[angle=0,width=\columnwidth]{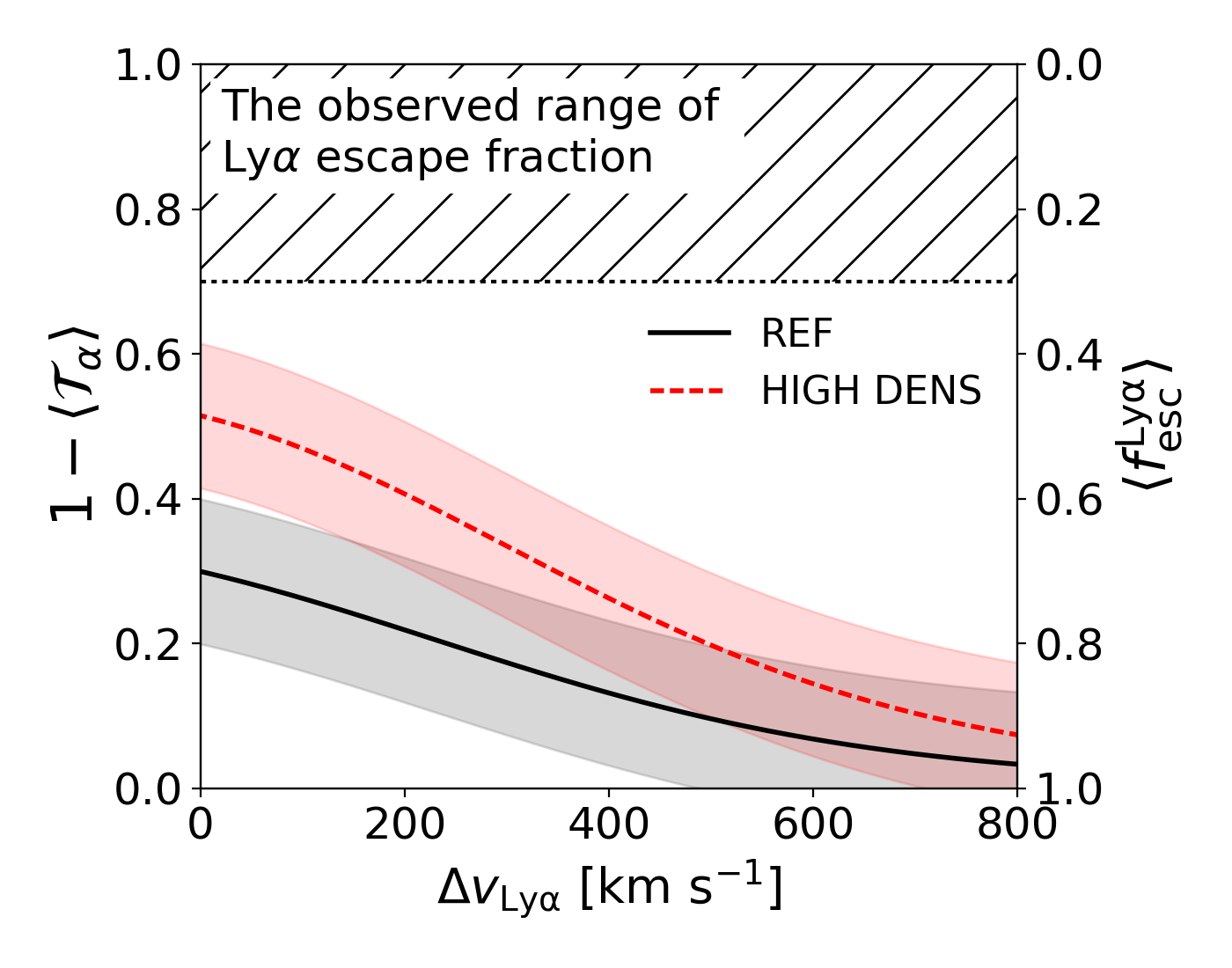}
  \vspace{-0.4cm}
  \caption{The contribution of the CGM and IGM on the Ly$\alpha$ escape fraction at $z\sim2-3$. The black solid (red dashed) curve shows the Ly$\alpha$ transmission of the CGM and IGM, $\langle\mathcal{T}_\alpha\rangle$, in the REF (HIGH DENS) model as a function of Ly$\alpha$ velocity offset. The grey and red shaded regions indicate the typical model uncertainty bracketing the line width between $100\rm~km~s^{-1}$ and $400\rm~km~s^{-1}$. The hatched region indicates the observed range of Ly$\alpha$ escape fraction derived from the Ly$\alpha$/H$\alpha$ ratio. The right y-axis shows the value of Ly$\alpha$ escape fraction if the CGM/IGM is only the source of opacity, i.e. $\langle f_{\rm esc}^{\rm Ly\alpha}\rangle=\langle T_\alpha\rangle$ The figure illustrates that to explain the observed Ly$\alpha$ escape fraction, galaxies should have a large contribution from the ISM, but there is a non-negligible impact from the CGM and IGM.}
  \label{fig:escape}
\end{figure}

We now examine the impact of the CGM and IGM on Ly$\alpha$ escape fraction. Observationally, $\langle f_{\rm esc}^{\rm Ly\alpha}\rangle$ can be estimated from galaxy spectra using the ratio between the observed Ly$\alpha$ line flux and the expected Ly$\alpha$ line flux inferred from the observed H$\alpha$ line flux, which are both produced following recombination with a known ratio. The Ly$\alpha$ flux is typically suppressed relative to this ratio, which reflects either that Ly$\alpha$ photons were efficiently destroyed by dust, or that Ly$\alpha$ photons were scattered into an extended low surface brightness halo. In practice, observations generally cannot distinguish between these two scenarios, while theoretically we expect both physical processes to affect the observationally inferred $ f_{\rm esc}^{\rm Ly\alpha}$.

Figure~\ref{fig:escape} shows the {\it contribution} of the CGM and IGM to the Ly$\alpha$ escape fraction in our model as a function of Ly$\alpha$ velocity offset from line center. The black solid (red dashed) curve shows the total Ly$\alpha$ escape fraction only taking into account the contribution from the CGM and IGM, i.e $\langle f_{\rm esc}^{\rm Ly\alpha}\rangle=\langle\mathcal{T}_\alpha\rangle$, in the REF model (HIGH DENS model). In Figure~\ref{fig:escape} also compares these curves to the typical observed range (hatched region) (e.g. \citealt{2010Natur.464..562H,2015ApJ...809...89T}). This shows that the contribution from the CGM and IGM alone is not sufficient to explain the observed Ly$\alpha$ escape fraction, {\it but the CGM has a non-negligible impact.} In addition, the contribution of the CGM to the Ly$\alpha$ escape fraction is larger at higher CGM densities (HIGH DENS model), which can introduce an environmental impact on the observationally inferred Ly$\alpha$ escape fraction. 

While the importance of the ISM controlling the Ly$\alpha$ escape is well appreciated in literature, our analysis provides independent support of this conclusion based on the galaxy-Ly$\alpha$ forest clustering data. However, our results imply that Ly$\alpha$ radiative transfer does not end after Ly$\alpha$ photons escape from galaxies, even at $z\sim 2-3$.

In detail, Ly$\alpha$ escape from the ISM and Ly$\alpha$ RT through the CGM/IGM are coupled: scattering in the ISM affects the velocity offset of Ly$\alpha$ from line center \citep{2006A&A...460..397V,2016ApJ...826...14G}. The observed range of  Ly$\alpha$ velocity offsets, $\Delta v_{\rm Ly\alpha}=0-600\rm~km~s^{-1}$ \citep{2010ApJ...717..289S} gives rise to IGM/CGM transmissions in the range $\langle \mathcal{T}_\alpha\rangle\approx0.70$ to $\approx0.90$ (REF model). Scattering through galactic outflows can give rise to large intrinsic Ly$\alpha$ velocity offsets, where the CGM and IGM are more transparent \citep[even during the reionization epoch, see][]{2010MNRAS.408..352D}. The environmental dependence of IGM/CGM transmission can have interesting implications for the large-scale clustering of LAEs and the possible non-gravitational contribution by the Ly$\alpha$ RT effect through the surrounding gas environments around Ly$\alpha$ emitting galaxies (see Sect.~\ref{sec:implication_cosmology}). 

\subsection{Ly$\alpha$ haloes}

\subsubsection{Self-consistent profile of Ly$\alpha$ haloes based on the galaxy-Ly$\alpha$ absorption clustering data}

\begin{figure*}
\centering
\includegraphics[angle=0,width=0.8\textwidth]{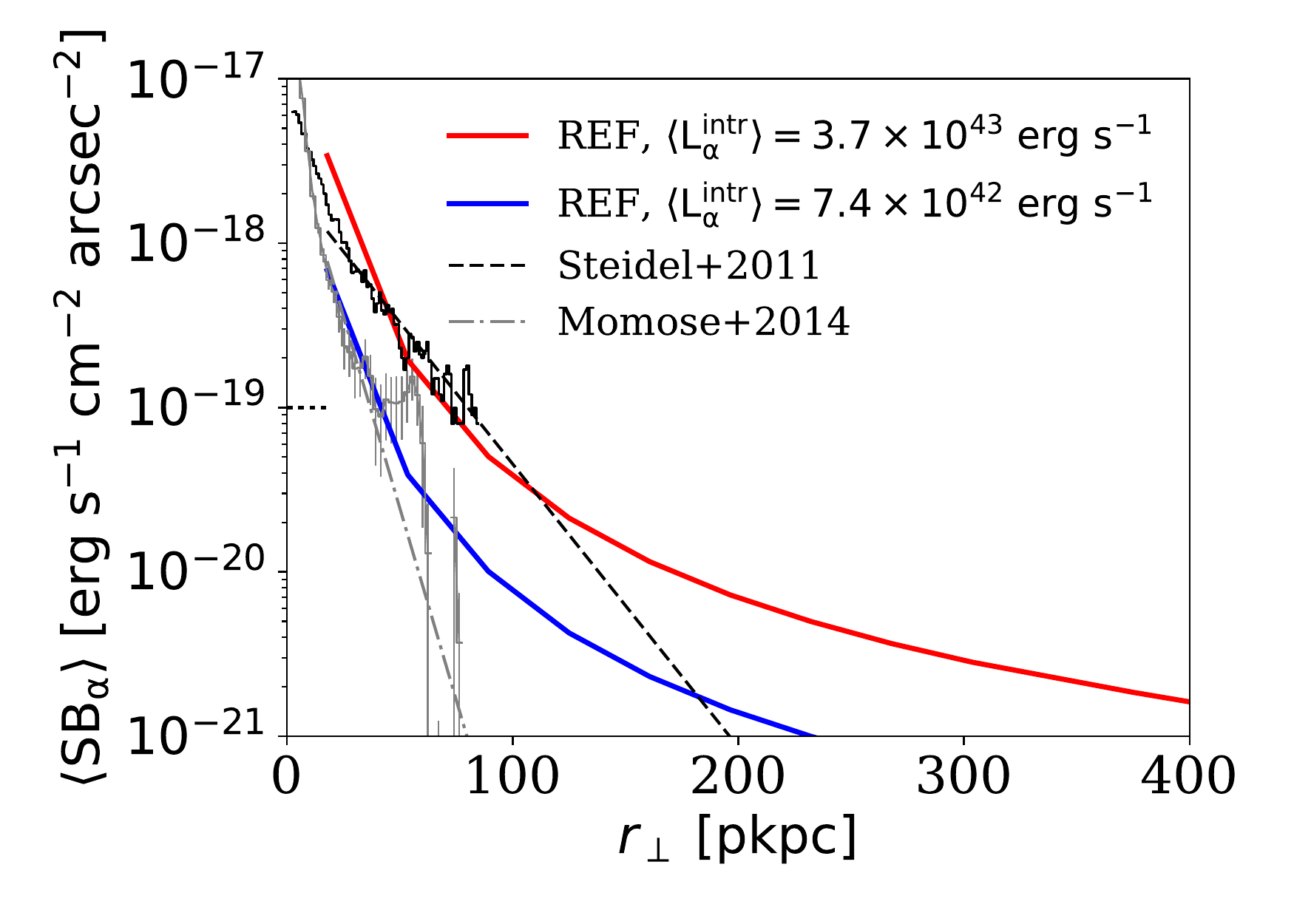}
\vspace{-0.4cm}
\caption{Mean surface brightness profile of Ly$\alpha$ haloes in the REF model with two different intrinsic Ly$\alpha$ luminosities $\langle L_\alpha^{\rm intr}\rangle=3.7\times10^{43}\rm~erg~s^{-1}$ (red curve) and $7.4\times10^{42}\rm~erg~s^{-1}$ (blue curve). The black and grey histograms show the observed surface brightness profiles of Ly$\alpha$ haloes around LBGs \citep{2011ApJ...736..160S} and LAEs \citep{2014MNRAS.442..110M}; the dashed and dash-dotted lines are their best-fit exponential profiles. The horizontal dotted line indicates the approximate surface brightness limit ($10^{-19}\rm~erg~s^{-1}~cm^{-2}~arcsec^{-2}$) of the observations.}\label{fig:halo}
\end{figure*}

Finally, we present our results for the Ly$\alpha$ haloes based on the galaxy-Ly$\alpha$ forest clustering data. Figure \ref{fig:halo} shows the model prediction of the mean surface brightness profile of Ly$\alpha$ haloes around star-forming galaxies. The red and blue curves show the Ly$\alpha$ haloes for two different intrinsic Ly$\alpha$ luminosities of the central galaxies, $\langle L_\alpha^{\rm intr}\rangle=3.7\times10^{43} $ and $7.4\times10^{42}\rm~erg~s^{-1}$. The former is our fiducial model with $\rm SFR=34\rm~M_{\odot}~yr^{-1}$. The latter shows a case for $\rm SFR=6.8\rm~M_{\odot}~yr^{-1}$, illustrating a dependence of Ly$\alpha$ haloes on the star formation activity of galaxies.

Comparing with the observed mean surface brightness profiles of LBGs (black histogram, \citealt{2011ApJ...736..160S}) and LAEs (grey histogram, \citealt{2014MNRAS.442..110M}), the model reproduces the mean surface brightness profile well in the observed range $r_\perp\approx20-80\rm~pkpc$. We emphasise that because the distribution and kinematics of the CGM and IGM are pre-constrained by the galaxy-Ly$\alpha$ forest clustering data (\S \ref{sec:calibration}), we did not adjust any parameter regarding the structure of the gaseous environment. At a fixed intrinsic Ly$\alpha$ luminosity, the model can be considered as a self-consistent and unique prediction of the mean surface brightness profile of Ly$\alpha$ haloes for a given formation mechanism (Ly$\alpha$ scattering of the central sources).

The clustering of neutral hydrogen gas around galaxies is responsible for producing the Ly$\alpha$ haloes. Ly$\alpha$ scattering in the mean IGM will underestimate the surface brightness of Ly$\alpha$ haloes. As the amount of neutral gas increases at inner radii, the intrinsic Ly$\alpha$ emission from the central galaxies are more likely to be scattered back onto the line of sight. While this decreases the visibility of the Ly$\alpha$ line in the galaxy spectra, these attenuated Ly$\alpha$ photons are not permanently lost. Some of the photons are scattered back toward observers and seen as the diffuse Ly$\alpha$ haloes around the galaxies. Furthermore, since we assume the powering by central star-forming galaxies, the normalisation of the surface brightness profile scales with the intrinsic Ly$\alpha$ luminosity, $\langle{\rm SB}_\alpha\rangle\propto\langle L_\alpha^{\rm intr}\rangle$.\footnote{Here, the intrinsic Ly$\alpha$ luminosity refers to the total amount of Ly$\alpha$ photons leaked out to the CGM and IGM from the ISM of galaxies. We note that the surprisingly good match to the observed Ly$\alpha$ haloes requires all Ly$\alpha$ photons produced within the ISM eventually leak out to the CGM. The effect of the absorption by dust enters as a lower value of the intrinsic Ly$\alpha$ luminosity. Therefore, we are not concluding that the Ly$\alpha$ scattering is the origin of Ly$\alpha$ haloes. Instead, we conclude that our joint Ly$\alpha$ emission - absorption modelling is consistent with the observation as the predicted contribution from Ly$\alpha$ scatterings is within the observed Ly$\alpha$ surface brightness. Nevertheless, despite the caveat, the remarkable match to the observation is worth noting.} The scaling with SFR can explain the difference between Ly$\alpha$ haloes around LBGs and LAEs. The surface brightness profile of Ly$\alpha$ haloes produced by scatterings of Ly$\alpha$ photons can be fitted by a power-law,
\begin{align}
&\left.\langle{\rm SB}_\alpha(r_\perp)\rangle\right/[\rm~erg~s^{-1}~cm^{-2}~arcsec^{-2}]\approx \nonumber \\
&~~~~~~~~~2.1\times10^{-18}\left(\frac{\langle L_\alpha^{\rm intr}\rangle}{3.7\times10^{43}{\rm~erg~s^{-1}}}\right)\left(\frac{r_\perp}{20\rm~ pkpc}\right)^{-2.4}.\label{eq:powerlaw}
\end{align}
This fit is accurate to $10$ per cent relative to the direct result from the model at $20{\rm~pkpc}<r_\perp<1000{\rm~pkpc}$.

The surface brightness profile also depends on the intrinsic Ly$\alpha$ velocity offset. When the intrinsic Ly$\alpha$ line profile has a smaller Ly$\alpha$ velocity offset, close to the line centre, the Ly$\alpha$ photons have more probability to be scattered by the surrounding CGM and IGM, increasing the surface brightness. However, this is a secondary effect compared to a dominating change caused by the intrinsic Ly$\alpha$ luminosity.

Over the radius 20-1000 pkpc, the mean surface brightness profile is well described by the power-law $\propto r^{\alpha}$ with the slope of  $\alpha\approx-2.4$. The power law tail extends out to $\sim1\rm~pMpc$ in the model prediction. This is in contrast with the conventional expontional profile $\propto \exp(-r/r_l)$ with a scale length $r_l$. Although within the observed range below $80\rm~pkpc$ both the power-law and exponential profiles describe the mean surface brightness profile of Ly$\alpha$ haloes, in the outskirt of the CGM, the exponential profile substantially underestimates the Ly$\alpha$ surface brightness.  Requiring the self-consistency with the galaxy-Ly$\alpha$ forest clustering data, the formation of Ly$\alpha$ haloes powered by the scattering of Ly$\alpha$ photons from central star-forming galaxies predicts that there must be large extent of diffuse Ly$\alpha$ emission from the CGM and IGM around galaxies. If this picture that Ly$\alpha$ haloes are powered by scattering is true, deeper observation should find an extended diffuse tail of Ly$\alpha$ haloes. 

We argue that the prediction of the extended power-law tail of Ly$\alpha$ emission in the diffuse haloes is a robust result {\it required} from the galaxy-Ly$\alpha$ absorption clustering data; the profile is difficult to change by changing the ISM and star forming properties of the galaxies such as the ISM Ly$\alpha$ escape fraction and SFR, which only enters as a rescaling of $\langle L_\alpha^{\rm intr}\rangle$. The prediction of a power-law profile of Ly$\alpha$ haloes is also supported from the Monte-Carlo Ly$\alpha$ RT calculation of the cosmological hydrodynamic simulation \citep{2017ApJ...835..207G}, which can also be fitted by a power-law profile with a similar slope.

Interestingly, the observation of the surface brightness profiles of Ly$\alpha$ haloes around QSOs by VLT/MUSE by \citet{2016ApJ...831...39B} reports a shallower power-law profile $\propto r^{-1.8}$ out to $\sim100\rm~pkpc$. This could suggest a large LyC leakage which lowers the surface brightness at inner radius or other Ly$\alpha$ emission mechanism such as fluorescence by the UV photons. Our constrained RT approach can be applied to understand the origin of the Ly$\alpha$ haloes of QSOs (e.g. Ly$\alpha$ scattering versus fluorescence) by extending a joint Ly$\alpha$ emission-absorption analysis \citep{2013ApJ...766...58H} using the QSO-Ly$\alpha$ forest clustering data \citep{2013ApJ...776..136P,2013JCAP...05..018F}.

The power-law emission tail of Ly$\alpha$ haloes may have an implication for the large-scale clustering of Ly$\alpha$ emission detected by \citet{2016MNRAS.457.3541C}, which will be discussed in \S \ref{sec:implication_cosmology}.

\begin{figure*}
\includegraphics[angle=0,width=\textwidth]{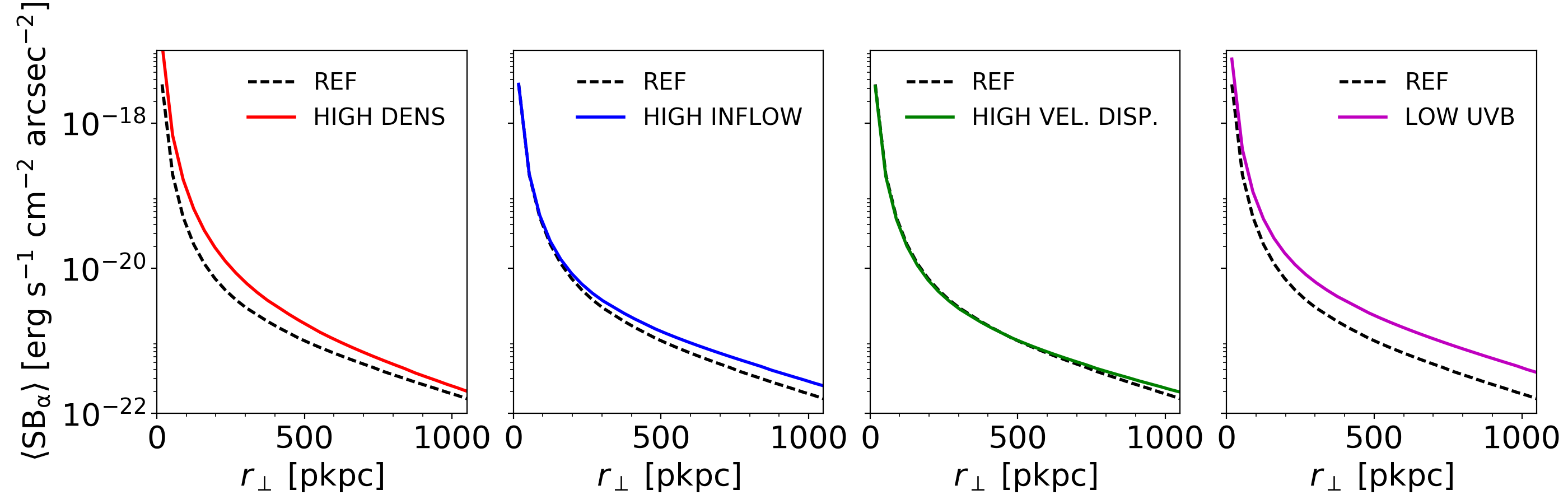}
\caption{Comparison of the mean surface brightness profiles of Ly$\alpha$ haloes in the REF model (black dashed curve) with HIGH DENS, HIGH INFLOW, HIGH VEL. DISP. and LOW UVB models (solid colored curves from left to right).The model parameters for REF model is shown in Table~\ref{table:model}. HIGH DENS model has $r_c=2h^{-1}\rm cMpc$, HIGH INFLOW model has $v_{\rm inflow}=300\rm~km~s^{-1}$, HIGH VEL. DISP. model has  $\sigma_v=300\rm~km~s^{-1}$, and LOW UVB model has $\Gamma_{\rm bkg}=2\times10^{-13}\rm~s^{-1}$. All other parameters are the same as REF model.}\label{fig:halo_MODELS}
\end{figure*}

\subsubsection{Dissecting the impacts of CGM/IGM: \newline the clustering and kinematics of the gas around galaxies}

Similar to emergent Ly$\alpha$ line profiles, we can examine the details of the impacts of the gas density, kinematics, and UV background of the CGM and IGM on the Ly$\alpha$ haloes. Figure \ref{fig:halo_MODELS} shows the different surface brightness profiles of Ly$\alpha$ haloes in different gaseous environments: a high density (HIGH DENS), a high inflow velocity (HIGH INFLOW), high velocity dispersion (HIGH VEL. DISP.), and low UV background (LOW UVB) models.

The clustering of neutral hydrogen gas around galaxies increases the surface brightness of the Ly$\alpha$ haloes. The increasing amount of gas increases the probability that Ly$\alpha$ photons are scattered back to lines of sight. For example, the high density environment with the correlation length of $r_c=2h^{-1}\rm cMpc$ increases the surface brightness profile by about a factor of two. This simultaneously increases the amount of Ly$\alpha$ absorption in that region probed by the background QSO spectra.

Unlike the Ly$\alpha$ line profile, the kinematics of the gas around galaxies only has a secondary role in determining the surface brightness profile of the Ly$\alpha$ haloes. This is because the surface brightness profile is integrated over the frequency, the velocity information is averaged over. There is nonetheless a small impact on the surface brightness profile. As the inflowing gas has more probability to scatter Ly$\alpha$ photons as the photons is redshifted into the line centre, higher inflow velocity can increase the surface brightness. Higher velocity dispersion similarly increases the probability that Ly$\alpha$ photons are scattered back to our line of sight at the outskirt of Ly$\alpha$ haloes, where the Ly$\alpha$ photons escape otherwise because of the dominating Hubble flow.

The photo-ionization rate of the UV background impacts the surface brightness of Ly$\alpha$ haloes. A lower photo-ionization rate increases the neutral fraction of gas, thus more Ly$\alpha$ photons can be scattered back into lines-of-sight to form larger Ly$\alpha$ haloes.

\section{Implications}\label{sec:implication}

\subsection{On the origin of Ly$\alpha$ escape and haloes: how do the CGM and IGM affect Ly$\alpha$ radiation in and around galaxies?}

\begin{figure}
\advance\leftskip-0.6cm
\includegraphics[angle=0,width=1.15\columnwidth]{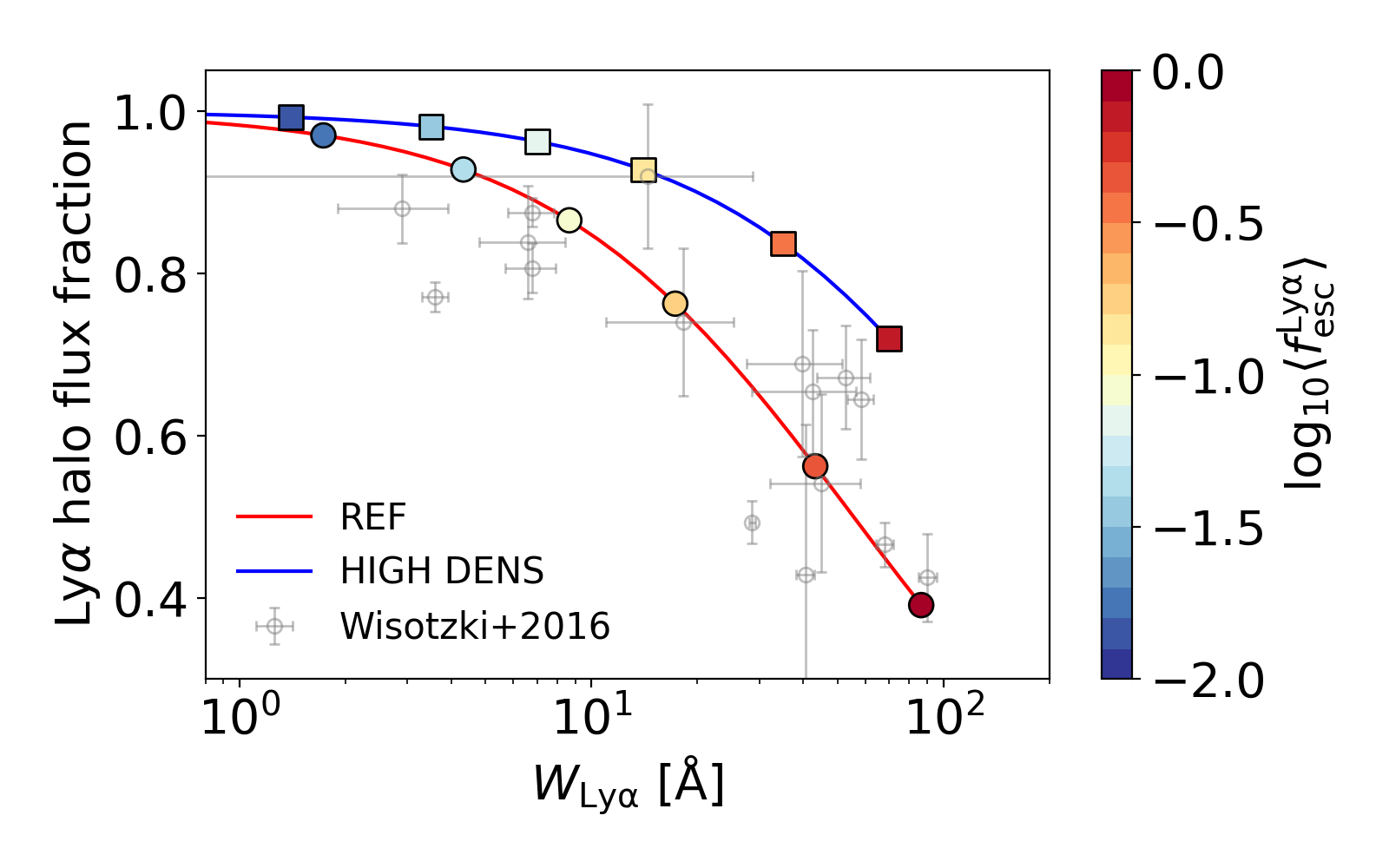}
\vspace{-0.6cm}
\caption{Ly$\alpha$ halo flux fraction, $X_{\rm Ly\alpha,halo}=F_{\rm halo}/(F_{\rm gal}+F_{\rm halo})$, as a function of of observed Ly$\alpha$ EW of the emission line of galaxies in the REF (red curve with filled circles) and HIGH DENS model (red curve with filled squares). The points corresponds to $\langle f_{\rm esc,ISM}^{\rm Ly\alpha}\rangle=0.02,0.05,0.10,0.20,0.50,1.0$ from left to right. The filled circles are colour-coded with the total Ly$\alpha$ escape fraction, $\langle f_{\rm esc}^{\rm Ly\alpha}\rangle=\langle f_{\rm esc,ISM}^{\rm Ly\alpha}\rangle\langle\mathcal{T}_\alpha\rangle$, as indicated by the colour bar. The data points from the MUSE {\it Hubble} Deep Field South by Wisotzki et al. (2016) are shown as open grey points.}\label{fig:halo_fraction}
\end{figure}

We discuss how the ISM, CGM, and IGM affect Ly$\alpha$ escape and Ly$\alpha$ haloes around galaxies. \citet{2016A&A...587A..98W} recently found an anti-correlation in the halo flux fraction - Ly$\alpha$ EW relation. In other words, larger Ly$\alpha$ halo flux fractions are generally found around galaxies that have a lower equivalent width of Ly$\alpha$ emission line emission coming {\it directly} from the galaxy. Figure \ref{fig:halo_fraction} compares our dust-free ISM + CGM/IGM models with varying ISM escape fractions with observations (see \S \ref{sec:halo_model} for a description of how we compute the halo flux fraction). Figure \ref{fig:halo_fraction} shows that:

\begin{itemize}
\item The agreement between the observed anti-correlation between Ly$\alpha$ halo flux fraction and Ly$\alpha$ EW and our model suggests that anti-correlation is indeed caused by the spatial `redistribution' of the galaxy's Ly$\alpha$ flux into extended Ly$\alpha$ haloes via scattering. Higher $\HI$ gas densities more easily scatter the nebular Ly$\alpha$ emission out of the line-of-sight of an observer, into a more diffuse Ly$\alpha$ fog. In this picture, varying gas density shifts the halo flux fraction - Ly$\alpha$ EW relation diagonally.  

Most Ly$\alpha$ RT occurs on interstellar scales (see \S~\ref{sec:escape}). If we invoke scattering as the main driver behind the observed anti-correlation between Ly$\alpha$ halo flux fraction and Ly$\alpha$ EW, then this must occur on interstellar scales. Note that `interstellar' scattering here refers to {\it all} scattering at $r < r_{\rm min}$. This interpretation sharply contrasts with a picture in which Ly$\alpha$ escape is dominated by dust. In the picture, dust does not only reduce the EW, but it also reduces the Ly$\alpha$ flux that escapes into the CGM, and therefore the surface brightness of Ly$\alpha$ haloes. This leads to an interesting implication: if interstellar RT dominates the Ly$\alpha$ scattering process (which it likely does), and if this scattering process provides the physical reason for the observed anti-correlation between Ly$\alpha$ halo flux fraction and Ly$\alpha$ EW, the role of dust on interstellar Ly$\alpha$ is weaker than expected. The $\HI$ gas density and kinematics of the ISM is a primary driver in regulating the Ly$\alpha$ escape.

\vspace{0.3cm}
\item Our analysis also shows that Ly$\alpha$ scattering in the CGM/IGM cannot be ignored, even at $z\sim 2-3$. We expect that the CGM and IGM to contribute to the scatter in the  Ly$\alpha$ halo flux fraction - Ly$\alpha$ EW relation (at fixed EW). A higher $\HI$ gas density in the CGM/IGM increases the surface brightness of Ly$\alpha$ haloes via scattering. Increasing $\HI$ gas density in the CGM/IGM enhances the Ly$\alpha$ halo flux fraction with a small suppression of Ly$\alpha$ EW. This leads to a (almost) vertical shift in the halo flux fraction - Ly$\alpha$ EW relation. 

For $W_{\rm Ly\alpha}\gtrsim20$~\AA~objects (which correspond approximately to narrow-band selected LAEs), the observed scatter in \citet{2016A&A...587A..98W} may be (partially) due to the environmental dependence of Ly$\alpha$ escape fraction and Ly$\alpha$ haloes introduced by the CGM and IGM. This interpretation is consistent with results of \citet{2012MNRAS.425..878M} -- the spatial extent of Ly$\alpha$ haloes around LAEs ($W_{\rm Ly\alpha}\gtrsim20$~\AA) depends on the Mpc-scale environment. 
\end{itemize}

The role of the CGM and IGM on the Ly$\alpha$ escape fraction and Ly$\alpha$ haloes through its environmental dependence can be tested observationally. If true, we should see a `thicker' Ly$\alpha$ forest around galaxies with brighter Ly$\alpha$ halos and a lower Ly$\alpha$ EW. The constrained RT approach provides us with quantitative predictions for the amount of Ly$\alpha$ absorption in the forest (e.g. Figure \ref{fig:HIGH_DENS_EW}). Such result can be tested with the MUSE QSO field data, where we expect both the measurements of individual Ly$\alpha$ haloes and the Ly$\alpha$ forest absorption in the vicinity of the Ly$\alpha$ emitting galaxies are possible. Alternatively, we could perform the Ly$\alpha$ forest tomography technique \citep{2014ApJ...795L..12L} in the MUSE {\it Hubble} Deep Field South field using bright background LBGs.

This discussion -- while speculative  -- nevertheless clearly demonstrates the effectiveness of a joint Ly$\alpha$ emission-absorption analysis using (integral field) spectroscopic survey of galaxies in QSO fields in shedding new light on the physical origin of Ly$\alpha$ escape and haloes.

\subsection{Cosmology with Ly$\alpha$ emitting galaxies}\label{sec:implication_cosmology}

\subsubsection{Diffuse Ly$\alpha$ emission in the Universe}

Our joint analysis with the galaxy-Ly$\alpha$ forest clustering data predicts the power-law emission tail of Ly$\alpha$ haloes. The Ly$\alpha$ haloes may thus contain more Ly$\alpha$ emission at larger scale than the exponential scale length, which is typically fit to the observations of Ly$\alpha$ haloes. The power-law surface brightness profile of Ly$\alpha$ haloes (equation \ref{eq:powerlaw}) contains $\sim1.5$ times more Ly$\alpha$ emission at radii larger than the exponential scale length ($25.2{\rm~pkpc}<r_\perp<5\rm~pMpc$) than the exponential profile of \citet{2011ApJ...736..160S}. Beyond the knee of the exponential profile, after matching the normalisation of the power-law profile at $60\rm~pkpc$ with the exponential profile, the power-law tail at $>60\rm~pkpc$ contains $\sim2-3$ times more Ly$\alpha$ emission than the exponential surface brightness profile of Ly$\alpha$ haloes. Therefore, the power-law emission tail of Ly$\alpha$ haloes contribute more to the diffuse Ly$\alpha$ emission of the Universe detected by \citet{2016MNRAS.457.3541C} than estimated using the conventional exponential profile. Although this boosts a contribution from Ly$\alpha$ haloes around star-forming galaxies to the diffuse Ly$\alpha$ emission background, the collective contribution from the Ly$\alpha$ emission from the power-law tails around galaxies still likely fall short below to explain the total Ly$\alpha$ emission of \citet{2016MNRAS.457.3541C}.

\subsubsection{Ly$\alpha$ RT effect on the large-scale LAE clustering}

The non-gravitational bias in the large-scale clustering of LAEs by the Ly$\alpha$ RT \citep{2011ApJ...726...38Z,2018A&A...614A..31B} may complicate a cosmological analysis in HETDEX. In this paper, we showed that the impact of the large-scale gaseous environments around galaxies on Ly$\alpha$ line flux can be estimated from galaxy-Ly$\alpha$ forest cross-correlation data. This means that if the IGM modulates the visibility of Ly$\alpha$ line of galaxies, and consequently the clustering signal of Ly$\alpha$-selected galaxies, we also expect a correlation between the Ly$\alpha$ EW of emission lines and galaxy-Ly$\alpha$ forest cross-correlation function. For example, the measurement of the LBG-Ly$\alpha$ forest cross-correlation function as a function of different  Ly$\alpha$ EW sub-samples of LBGs could test the impact of Ly$\alpha$ RT effect on the large-scale clustering of Ly$\alpha$ emitting galaxies. 

The large-scale UV background fluctuations, which may also affect the Ly$\alpha$ forest auto-correlation function \citep{2014PhRvD..89h3010P,2014MNRAS.442..187G}, can perturb the visibility of Ly$\alpha$ lines and would induce an additional non-gravitational contribution \citep{2011MNRAS.415.3929W}. To circumvent these issues, various joint HETDEX+BOSS analysis among Ly$\alpha$ line profile, galaxy-Ly$\alpha$ forest cross-correlation, auto-correlation function of Ly$\alpha$ emitting galaxies, and Ly$\alpha$ forest auto-correlation function, are important to quantify the astrophysical impact of the CGM/IGM on Ly$\alpha$ line of galaxies, and hence isolate the cosmological contribution in the large-scale LAE clustering from non-gravitational ones. A further study of Ly$\alpha$ RT modelling and Ly$\alpha$ forest using cosmological hydrodynamical simulations of galaxies and the IGM is required to address this issue more carefully.

\subsection{Caveats}

The analysis presented in this paper is at an early stage. Ideally the same galaxy population should be selected for all observations: galaxy-Ly$\alpha$ forest clustering, Ly$\alpha$ escape fraction, and Ly$\alpha$ haloes. The present analysis relied on LBG sample for galaxy-Ly$\alpha$ forest clustering, which has then been assumed to be the same for all galaxies that escape fraction and Ly$\alpha$ haloes are measured from. This clearly introduces uncertainties in our results. For example, for LAEs because the gas overdensity around lower mass haloes is expected to be lower, the surface brightness of the Ly$\alpha$ haloes may be reduced as less photons are scattered back into lines of sight. While we have explored the possible variation by varying the model parameters around the calibrated value, future analysis will benefit from the measurements of all the three observables from the uniformly selected sample. 

While our main approach -- constrained Ly$\alpha$ radiative transfer -- can substantially reduce the number of free parameters, a particular model adopted in this paper still contains some arbitrariness. Some of which, e.g. gas velocity field, can be improved by directly fitting the model to the data by the Markov chain Monte Carlo method. We have for example neglected the impact of outflow. The outflow can allow more Ly$\alpha$ photons to escape and reduce the surface brightness of Ly$\alpha$ haloes at smaller radii. Although the currently available 2D galaxy-Ly$\alpha$ forest clustering measurement does not show clear outflow signature, once the measurement is improved the model should be generalised to allow the effect of outflow. Furthermore, the analysis will benefit by replacing the fundamental assumption of the model, i.e. single scattering approximation, with a full Monte-Carlo Ly$\alpha$ radiative transfer calculation. Although the single scattering approximation seems to match the full Monte-Carlo result reasonably well \citep{2012MNRAS.424.1672D}, the full calculation eliminates the need of somewhat arbitrary choice of $r_{\rm min}$ in the model (see Appendix \ref{app:innermost} for the assessment of the uncertainty).

\section{Conclusions}\label{sec:conclusion}

We have presented a new approach to ``constrained Ly$\alpha$ radiative transfer'' through the CGM and IGM based on a perturbative expansion of the scattering process (\S~\ref{sec:methodology} and \ref{sec:theory}). The central idea is to perform Ly$\alpha$ RT calculations through a CGM and IGM whose $\HI$ content and kinematics are constrained by the observed cross-correlation between galaxies and Ly$\alpha$ forest absorption lines (see \S \ref{sec:calibration}). This enables us to quantify how Ly$\alpha$ propagates around galaxies in realistic, observationally-constrained, cosmological environments. We apply this approach to investigate how Ly$\alpha$ escapes from galaxies and their environment. We specifically focused on studying ({\it i}) how the CGM/IGM impacts the Ly$\alpha$ line flux, and spectral line profiles that we observe directly from galaxies, and ({\it ii}) the mean surface brightness profile of extended Ly$\alpha$ emission produced by scattering in the CGM/IGM. Our analysis, which focussed on $z\sim 2-3$, showed that \\

\begin{itemize}
\item The CGM and IGM at $z\sim 2-3$ transmit $\langle\mathcal{T}_\alpha\rangle\approx80~\%$ of Ly$\alpha$ line emission escaping from galaxies at a Ly$\alpha$ velocity offset of $\Delta v_{\rm Ly\alpha}\approx300\rm~km~s^{-1}$ (redward of the line centre). The transmission varies from $ \approx60~\%$ at $\Delta v_{\rm Ly\alpha}\approx100\rm~km~s^{-1}$ to $\approx90~\%$ at $\Delta v_{\rm Ly\alpha}\approx600\rm~km~s^{-1}$. This excess attenuation of Ly$\alpha$ photons by the CGM/IGM is due to the large overdensity of CGM gas relative to that of mean IGM. The wavelength-dependence of the IGM transmission curve near the systemic velocity of a galaxy is determined by the gas kinematics of the CGM. Comparing with observational constraints on the Ly$\alpha$ escape fraction at these redshifts (derived for example from Ly$\alpha$/H$\alpha$ ratios, e.g. \citealt{2010Natur.464..562H}), our model implies that the ISM plays the biggest role in setting the Ly$\alpha$ escape fraction. This confirms the traditional view that the ISM is a primary driver of the Ly$\alpha$ escape. However, { \it our results demonstrate that there is a non-negligible impact of the CGM and IGM on the Ly$\alpha$ line even at $z\approx2-3$.}\\

\item We show that Ly$\alpha$ scattering in the CGM gives rise to Ly$\alpha$ haloes with a power-law emission tail $\propto r^{-2.4}$ at $\gtrsim80\rm~pkpc$ extending out to the outskirts of the CGM and beyond ($\gtrsim300~\rm pkpc$, see equation~\ref{eq:powerlaw}). This result is robust, and does not depend on interstellar RT effects. This power-law profile differs from the often-assumed exponential profile, for which the the surface brightness drops rapidly beyond the exponential scale length. The extended power-law tail is a result of the extended clustering of cold gas around galaxies, that is required by the galaxy-Ly$\alpha$ forest clustering data. This gas scatters Ly$\alpha$ photons back into our line of sight to form Ly$\alpha$ haloes. Deeper observations of Ly$\alpha$ haloes probing the outskirts of the CGM will test this picture. Moreover, if we assume that {\it all} Ly$\alpha$ photons produced in the ISM eventually leak out into the CGM and IGM (note that `ISM' here refers to $r< r_{\rm min}$), then Ly$\alpha$ scattering alone can reproduce the observed surface brightness as well as the observed anti-correlation between Ly$\alpha$ halo flux fraction and Ly$\alpha$ EW (see \S~\ref{sec:implication} and Figure~\ref{fig:halo_fraction}), surprisingly well. Of course, the 100 per cent Ly$\alpha$ leakage into the CGM is at face value unrealistic. However, the surprisingly good match may still imply that Ly$\alpha$ scatters abundantly at $r < r_{\rm min}$ without being efficiently destroyed by dust (see \S~\ref{sec:implication}), which places interesting constraints on the $\HI$ and dust distribution in the ISM (see e.g. \citealt{Gronke17}). \\

\item  The CGM and IGM cause a preferential suppression of the Ly$\alpha$ flux - and therefore by extension Ly$\alpha$ EW - of galaxies, and an overall enhancement of the surface brightness of Ly$\alpha$ haloes around galaxies that reside in denser parts of the Universe (more precisely,  inside cosmic volumes with a higher Ly$\alpha$ forest opacity in the background QSO spectra). The impact of the CGM and IGM can then introduce a large-scale environmental dependence of Ly$\alpha$ line profiles and Ly$\alpha$ haloes of galaxies. Studies of the environmental impact on Ly$\alpha$ line profiles and Ly$\alpha$ haloes, i.e. Ly$\alpha$ emission properties as a function of the large-scale gaseous environment, are important to fully understand the escape mechanism of Ly$\alpha$ photons and the physical origin of Ly$\alpha$ haloes. 
\end{itemize}

Finally, we would like to stress the tremendous potential of the joint Ly$\alpha$ emission - absorption measurements for studying the physics of CGM, reionization, and cosmology:\\
\begin{itemize}
\item From the theoretical side, the analysis introduced in this paper can easily be expanded in various ways. For example, one can study the redshift evolution of $\langle \mathcal{T}_{\alpha} \rangle$, Ly$\alpha$ haloes, and the UV background self-consistently to address the impact of the CGM and UV background on the decline of observed Ly$\alpha$ emission line at $z>6$ to study the reionization process (\citealt{2017ApJ...839...44S}). Furthermore, the model can easily be extended to e.g. ({\it i}) predict the spectra of the spatially scattered Ly$\alpha$ halos, and ({\it ii}) to include the contribution of fluorescence radiation around galaxies and QSOs (e.g. \citealt{2005ApJ...628...61C,2010ApJ...708.1048K,2016ApJ...822...84M}). \\

\item From the observational side, wide field imaging campaigns such as the Dark Energy Survey (DES) and Hyper Suprime-Cam Subaru Strategic Program (HSC-SSP) will provide new high-redshift QSO target fields. By combining these observations with data obtained with modern integral field spectrographs such as VLT/MUSE and Keck/KCWI (and in the near future JWST and ground-based 30 m telescopes), it becomes possible to provide a cosmic map of both Ly$\alpha$ emission and absorption. HETDEX galaxies and BOSS Ly$\alpha$ forests will also offer a promising dataset. Such maps provide invaluable probes of the detailed galaxy properties and the physical state of the CGM and IGM at $2\lesssim z\lesssim 7$. This approach will shed new light on the physics of CGM, hydrogen and helium reionization, and cosmology. 

\end{itemize}

\section*{Acknowledgments}

We thank Max Gronke and Mike Anderson for enlightening discussion; Kristian Finlator for pointing out the importance of Ly$\alpha$ haloes, which motivated us to examine the properties of Ly$\alpha$ haloes in this paper; Rieko Momose for providing us the tabulated surface brightness profile of Ly$\alpha$ haloes; Andrew Pontzen, Andreu Font-Ribera, and Richard Ellis for reading and the comments on the manuscript. We thank the referee for carefully reading the manuscript and her/his constructive comments. KK acknowledges support from the European Research Council Advanced Grant FP7/669253.

\bibliographystyle{mnras}
\bibliography{Reference}

\appendix

\section{Derivation of Ly$\alpha$ halo surface brightness}\label{sec:A2}

We describe a heuristic derivation of the mean surface brightness of Ly$\alpha$ haloes powered by the scattering of Ly$\alpha$ photons from central galaxies. The total bolometric Ly$\alpha$ luminosity within a comoving radius $R$ is 
\begin{equation}
\langle L_\alpha(<R)\rangle=\int_0^{R} dr\iint L_\alpha^{\rm abs}(r,v_r,\NHI)f(r,v_r,\NHI)dv_rd\NHI,
\end{equation}
where $L_\alpha^{\rm abs}(r,v_r,\NHI)$ is a Ly$\alpha$ luminosity of an individual absorber at a comoving distance $r$ away from the central galaxy, moving with a radial peculiar velocity $v_r$, with a $\HI$ column density $\NHI$, and $f(r,v_r,\NHI)drdv_rd\NHI$ is the phase-space distribution function of absorbers around galaxies, i.e. the expected number of absorbers within a phase space volume $(r,r+dr)$, $(v_r,v_r+dv_r)$, and $(\NHI,\NHI+d\NHI)$. The phase-space distribution function can be decomposed into a product of a real-space distribution function, $f_{\rm r}(r,\NHI)dr\NHI$, and the conditional probability distribution function of the peculiar velocities of absorbers at a given $r$, $p_{\rm v}(v_r|r)dv_r$, leading $f(r,v_r,\NHI)=f_{\rm r}(r,\NHI)p_{\rm v}(v_r|r)$.

The real-space distribution function is simply given by
\begin{equation}
f_{\rm r}(r,\NHI)dr\NHI=\frac{dn_{\rm abs}}{d\NHI}[1+\xi(r)]4\pi r^2drd\NHI,
\end{equation}
where $\frac{dn_{\rm abs}}{d\NHI}d\NHI$ is the comoving number density of absorbers in a range between $\NHI$ and $\NHI+d\NHI$ and $\xi(r)$ is the real-space correlation function between galaxies and absorbers. Note that $\frac{dn_{\rm abs}}{d\NHI}d\NHI$ is related to the $\HI$ density distribution function $\CDDF$ as \citep{2005ARA&A..43..861W}
\begin{equation}
\CDDF d\NHI dz=(1+z)^3\frac{dn_{\rm abs}}{d\NHI}\sigma_{\rm abs}\left|\frac{dl_p}{dz}\right|d\NHI dz.
\end{equation}

Furthermore, for the Gaussian streaming model the conditional probability distribution function of the peculiar velocity of absorbers around galaxies is
\begin{equation}
p_{\rm v}(v_r|r)=\frac{1}{\sqrt{2\pi \sigma_v^2(r)}}\exp\left[-\frac{(v_r-\langle v_r(r)\rangle)^2}{2\sigma_r^2(r)}\right].
\end{equation}

Therefore, using equation (\ref{eq:absorber_luminosity}) for the Ly$\alpha$ luminosity of individual absorbers, after some algebra, we obtain
\begin{align}
&\langle L_\alpha(<R)\rangle=\int_0^R dr\int d\NHI\CDDF\left|\frac{dz}{dr}\right|\times \nonumber \\
&\int\frac{dv_r}{\sqrt{2\pi\sigma_v^2(r)}}\mathcal{L}_\alpha(r,v_r,\NHI)\left[1+\xi(r)\right]\exp\left[-\frac{(v_r-\langle v_r(r)\rangle)^2}{2\sigma_v^2(r)}\right],
\end{align}
where we simplied the expression by introducing an auxiliary quantity,
\begin{equation}
\mathcal{L}_\alpha(r,v_r,\NHI)=\int \left[1-e^{-\tau_{\rm a}(\nu_{\rm inj},\NHI)}\right]L_\nu^{\rm intr}(\nu_e)d\nu_e.
\end{equation}

As the comoving bolometric emissivity (luminosity density) (in units of $\rm erg~s^{-1}~cMpc^{-3}$) is related to total luminosity within comoving radius $R$, $\int_0^R\langle\varepsilon_\alpha(\nu,r)\rangle 4\pi r^2dr=\langle L_\alpha(<R)\rangle$, we have $dL_\alpha(<R)/dR=\langle\varepsilon_\alpha(R)\rangle 4\pi R^2$. Thus, the $\LyA$ emissivity by the scattered radiation is given by
\begin{align}
&\langle\varepsilon_\alpha(r)\rangle=\frac{1}{4\pi r^2}\int d\NHI\CDDF\left|\frac{dz}{dr}\right|\times \nonumber \\
&\int\frac{dv_r}{\sqrt{2\pi\sigma_v^2(r)}}\mathcal{L}_\alpha(r,v_r,\NHI)\left[1+\xi(r)\right]\exp\left[-\frac{(v_r-\langle v_r(r)\rangle)^2}{2\sigma_v^2(r)}\right],
\end{align}
which conclude the derivation of equation (\ref{eq:Lya_emissivity}). The mean surface brightness of Ly$\alpha$ haloes follows immediately by integrating the Ly$\alpha$ emissivity along a line-of-sight at each impact parameter.

\section{Impact of photoionization by local ionizing sources}\label{app:CDDF}

By taking the Jeans argument by \cite{2001ApJ...559..507S}, the $\HI$ column density of an absorber scales as $\NHI\propto \Gamma^{-1}$ as
\begin{align}
&\NHI\sim 2.7\times10^{13}{\rm{cm^2}}(1+\delta)^{3/2}\left(\frac{T}{10^4{\rm~K}}\right)^{-0.26}\left(\frac{\Gamma}{10^{-12}{\rm~s^{-1}}}\right)^{-1}
\nonumber \\ 
&\times\left(\frac{1+z}{4}\right)^{9/2}
\left(\frac{\Omega_bh^2}{0.02}\right)^{3/2}\left(\frac{f_g}{0.16}\right)^{1/2}.
\end{align}
This means that at a fixed density fluctuation $\delta$, different photoionization rates give rise to different $\HI$ column densities of absorbers. Therefore, the impact of photoionization can be taken into account by rescaling the value of $\HI$ column density at a photoionization rate $\Gamma_{\rm bkg}$ to a new value of photoionization rate $\Gamma(r)=\Gamma_{\rm local}(r)+\Gamma_{\rm bkg}$ which includes the photionization rate from local ionizing sources $\Gamma_{\rm local}(r)$. The $\HI$ column density of absorbers in the vicinity of galaxies is given by rescaling the column density of absorbers in the average IGM $\NHI^{\rm{bkg}}$, \begin{equation}
\NHI=\left[\frac{\Gamma_{{\rm bkg}}}{\Gamma(r)}\right]\NHI^{{\rm{bkg}}}.
\end{equation}

Following \cite{1997ApJ...486..599H}, the CDDF around galaxies including the photoionization rate of the local sources is given by rescaling the CDDF at the average IGM,  
\begin{equation}
\frac{\partial^2\mathcal{N}(r)}{\partial\NHI\partial z}=
\frac{\Gamma(r)}{\Gamma_{\rm{bkg}}}\left.
\frac{\partial^2\mathcal{N}}{\partial\NHI\partial z}\right|_{\rm bkg}.
\end{equation}
While here we explicitly indicated $\left.\frac{\partial^2\mathcal{N}}{\partial\NHI\partial z}\right|_{\rm bkg}$ as the CDDF at the average UV background, the subscript $\rm bkg$ is dropped in the main text. Approximating the CDDF with a power-law,
\begin{equation}
\left.
\frac{\partial^2\mathcal{N}}{\partial\NHI\partial z}\right|_{\rm bkg}
=A(z)(\NHI^{\rm{bkg}})^{-\beta_{\rm eff}},
\end{equation}
where $A(z)$ is a constant of proportionality, we obtain the CDDF around galaxies as
\begin{align}
\frac{\partial^2\mathcal{N}(r)}{\partial\NHI\partial z}
&=A(z)\left[\frac{\Gamma(r)}{\Gamma^{\rm{bkg}}}\right]^{-\beta_{\rm eff}+1}\NHI^{-\beta_{\rm eff}} \nonumber \\
&=\left[\frac{\Gamma(r)}{\Gamma_{\rm{bkg}}}\right]^{-\beta_{\rm eff}+1}\left.
\frac{\partial^2\mathcal{N}}{\partial\NHI\partial z}\right|_{\rm bkg}.
\end{align}
The factor $\left[\frac{\Gamma(r)}{\Gamma_{\rm{bkg}}}\right]^{-\beta_{\rm eff}+1}$ indicates the radial suppression of the normalization of the CDDF due to the photoionization by local ionizing sources. Therefore, by comparing with the definition of the real-space correlation function, we can identify the photoionization correction factor $C^{\rm phot}(r)$ as 
\begin{equation}
C^{\rm phot}(r)=\left[\frac{\Gamma(r)}{\Gamma_{\rm{bkg}}}\right]^{-\beta_{\rm eff}+1}=
\left[\frac{\Gamma_{\rm local}(r)}{\Gamma_{\rm{bkg}}}+1\right]^{-\beta_{\rm eff}+1}.
\end{equation}
By substituting $\displaystyle\Gamma_{\rm local}(r)=\int_{\nu_{912}}^\infty  \sigma_{\mbox{\tiny{HI}}}(\nu)\frac{L_\nu(\nu)}{4\pi [r/(1+z)]^2}\frac{d\nu}{h\nu}$, after some algebra we arrive at equations (\ref{eq:correction_factor1}) and (\ref{eq:correction_factor2}).

\section{The innermost radius}\label{app:innermost}

\begin{figure}
 \begin{center}
  \includegraphics[angle=0,width=0.8\columnwidth]{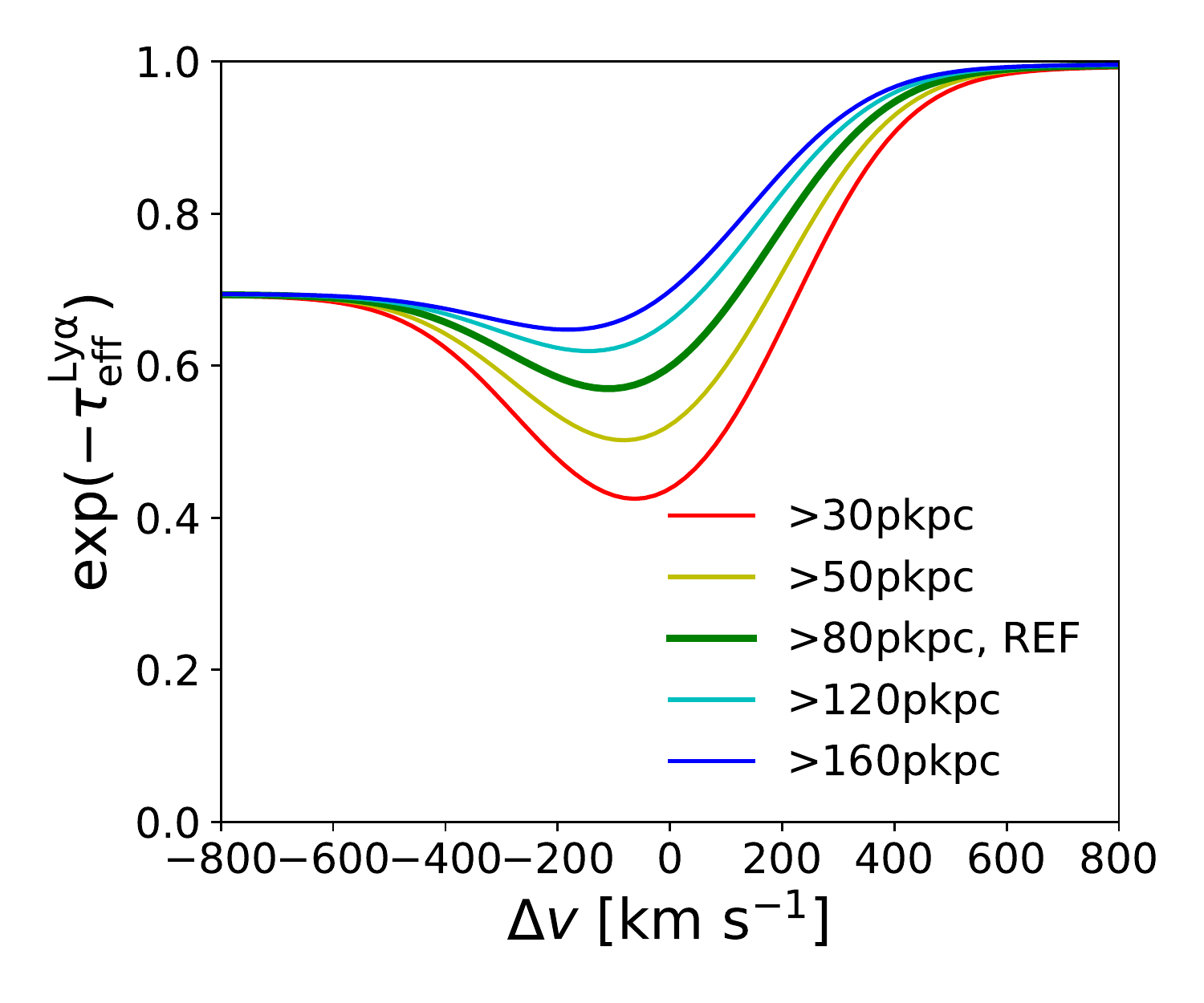}
  \caption{Dependence of the CGM/IGM transmission curve on the innermost radius.}
   \label{fig:innermost}
 \end{center}
\end{figure}

\begin{figure*}
 \begin{center}
  \includegraphics[angle=0,width=0.8\columnwidth]{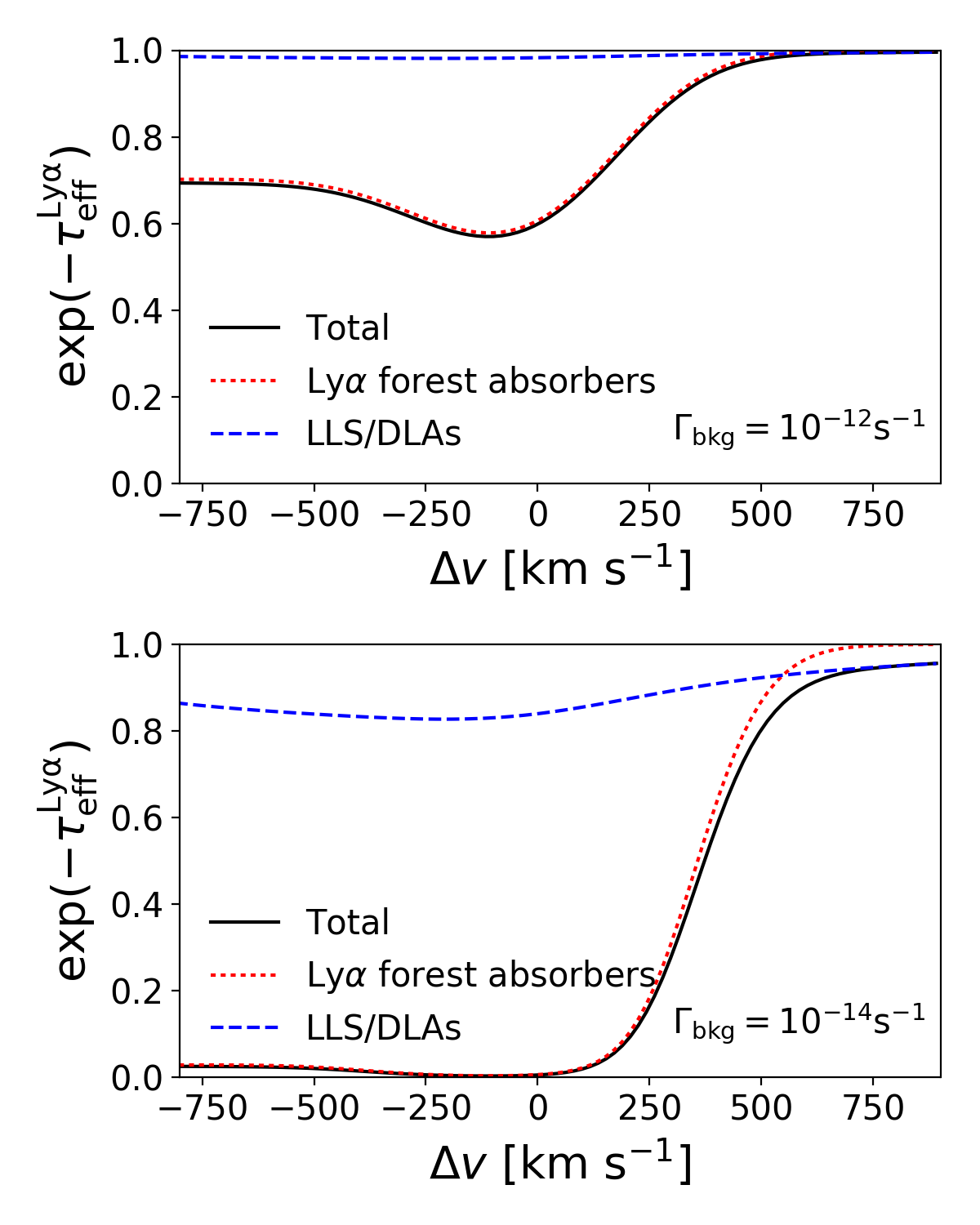}
  \hspace*{-2mm}
   \includegraphics[angle=0,width=0.8\columnwidth]{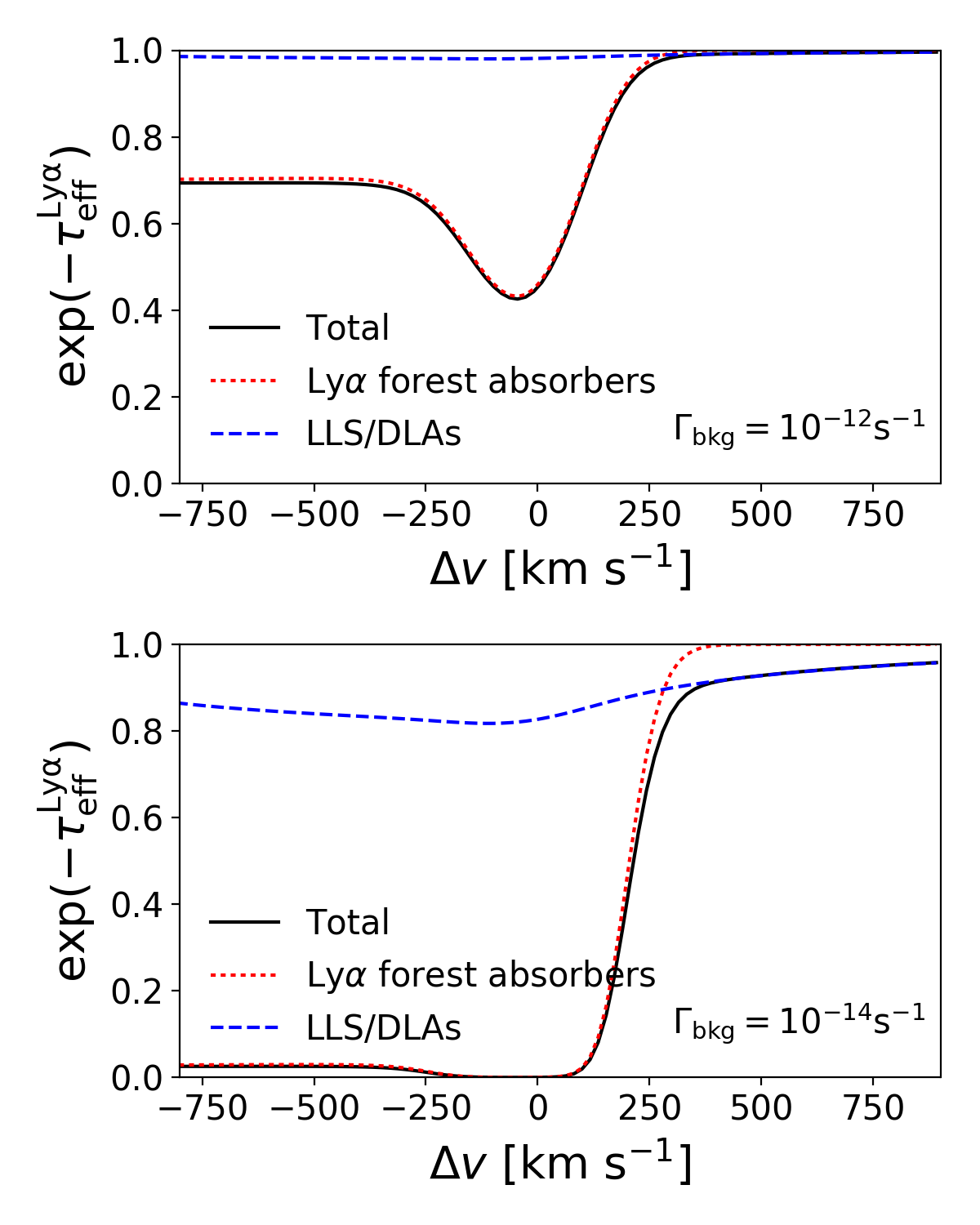}
  \caption{The CGM/IGM transmission curves including all absorbers (black solid, $10^{12}<\NHI/\rm cm^{-2}<10^{21.55}$), only Ly$\alpha$ forest absorbers (red dotted, $10^{12}<\NHI/\rm cm^{-2}<10^{17}$), and only LLS and DLAs (blue dashed, $10^{17}<\NHI/\rm cm^{-2}<10^{21.55}$). {\bf Left panels}: Models with the velocity dispersion $\sigma_v=200\rm~km~s^{-1}$ and $\Gamma_{\rm bkg}=10^{-12}\rm~s^{-1}$ (top) or $\Gamma_{\rm bkg}=10^{-14}\rm~s^{-1}$ (bottom). {\bf Right panels}: Model with  the velocity dispersion $\sigma_v=100\rm~km~s^{-1}$ and $\Gamma_{\rm bkg}=10^{-12}\rm~s^{-1}$ (top) or $\Gamma_{\rm bkg}=10^{-14}\rm~s^{-1}$ (bottom). All the other parameters are the same as REF model. The figure shows the contribution of the different types of absorbers to the CGM/IGM transmission curves.}
   \label{fig:NHI}
 \end{center}
\end{figure*}

Figure~\ref{fig:innermost} shows the dependence of the Ly$\alpha$ visibility on the innermost radius $r_{\rm min}$ in the REF model, where $r_{\rm min}=30,~50,~80,~120,~160\rm~pkpc$ (red, yellow, green, cyan, blue curves). The virial radius of $9\times10^{11}\rm~M_\odot$ halo at $z=3$ is $r_{\rm vir}\approx80\rm~pkpc$. The opacity increases with a smaller innermost radius. Our fiducial choice adopted in this paper is $r_{\rm min}=r_{\rm vir}$. The difference between the recommendation of \citet{2011ApJ...728...52L}, $r_{\rm min}=1.5r_{\rm vir}$, with our fiducial choice is about 20 per cent. Using our fiducial ISM Ly$\alpha$ line profile, the Ly$\alpha$ transmission is $\langle \mathcal{T}_\alpha\rangle=0.76,~0.80,~0.84,~0.86,~0.88$ for $r_{\rm min}=30,~50,~80,~120,~160~\rm pkpc$. The extra scattering between the viral radius of halo ($80\rm~pkpc$) to the inner region of the model CGM ($\rm30~pkpc$) is $\langle \mathcal{T}_\alpha(>80\rm~pkpc)\rangle-\langle \mathcal{T}_\alpha(>30\rm~pkpc)\rangle=0.08$. Therefore, the estimated Ly$\alpha$ transmission is therefore robust against the choice of the innermost radius at about 10 per cent level. 

Decreasing the innermost radius to $r_{\rm min}=30\rm~pkpc$ increases the opacity by $\sim30$ per cent relative to $r_{\rm min}=r_{\rm vir}$. \citet{2011ApJ...728...52L} showed that below a virial radius the multiple scatterings affects the formation of Ly$\alpha$ lines; hence, $e^{-\tau}$ approximation becomes increasingly inaccurate. On this scale, a Ly$\alpha$ RT simulation is required to estimate the impact of the gas around galaxies on the Ly$\alpha$ line profile. As a compromise between the recommendation of \citet{2011ApJ...728...52L} and the smallest radius probed by Ly$\alpha$ absorption using galaxy-galaxy and galaxy-QSO pairs (below this scale, the properties of gas around galaxies are not well constrained), we choose $r_{\rm min}=80\rm~pkpc$ to evaluate the impact of the intergalactic environment `constrained by Ly$\alpha$ absorption' on the Ly$\alpha$ flux from Ly$\alpha$ emitting galaxies.

\section{Relative contribution of absorbers}\label{app:absorbers}

Figure~\ref{fig:NHI} shows the contribution of different absorbers and different velocity dispersion to the red damping wing. At $\Gamma_{\rm bkg}=10^{-12}\rm~s^{-1}$ the velocity dispersion of infalling low column density Ly$\alpha$ forest absorbers dominate the formation of red damping wing. The smoothness of the damping wing depends on the velocity dispersion of the CGM: a higher $\sigma_v$ smear out the damping wing. Note that the velocity dispersion parameter is the most poorly constrained parameter in our model due to the lack of tabulated 2D effective optical depth map. For example, decreasing to $\sigma_v=100\rm~km~s^{-1}$ the damping wing extends only to $\Delta v\approx250\rm~km~s^{-1}$, more consistent with the results of \citet{2011ApJ...728...52L} (see also \citet{2007MNRAS.377.1175D} where the effect of velocity dispersion is ignored).

Our finding is consistent with \citet{2016MNRAS.463.4019K}, where we find that at $z\sim7$ LLS and DLAs dominate the contribution to the red damping wing due to small-scale absorbers (e.g. W2 model). By taking a value of $\Gamma_{\rm bkg}=10^{-14}\rm~s^{-1}$, Figure~\ref{fig:NHI} indeed shows that the contribution from LLSs and DLAs dominate the red damping wing opacity at $\Delta v>250-500\rm~km~s^{-1}$, while it is negligible at $\Gamma_{\rm bkg}=10^{-12}\rm~s^{-1}$. This is because at $\Gamma_{\rm bkg}=10^{-12}\rm~s^{-1}$, although the effective optical depth at large $\Delta v$ is dominated by LLS/DLAs, its absolute value is very small, $\tau_{\rm eff}^{\rm Ly\alpha}\ll 1$; hence, it is negligible in $\exp(-\tau_{\rm eff}^{\rm Ly\alpha})$. As $\tau_{\rm eff}^{\rm Ly\alpha}\propto\Gamma_{\rm bkg}^{-1/2}$, at a lower photoionization rate $\Gamma_{\rm bkg}=10^{-14}\rm~s^{-1}$, the effective optical depth becomes $\tau_{\rm eff}^{\rm Ly\alpha}\gtrsim\mathcal{O}(1)$, making an appreciable contribution to $\exp(-\tau_{\rm eff}^{\rm Ly\alpha})$.

\label{lastpage}

\end{document}